\documentclass[twocolumn,english,prx]{revtex4-1}
\usepackage{color}
\usepackage[utf8]{inputenc}
\usepackage{graphicx}
\usepackage[T1]{fontenc}
\usepackage{float}
\usepackage{natbib}
\usepackage{textcomp}
\usepackage{amsmath}
\usepackage{amssymb}
\usepackage{graphicx}
\usepackage{esint}
\usepackage{natbib}
\usepackage{xcolor}
\usepackage{multirow}
\usepackage{ulem}

\begin{document}
\title{Aharonov-Bohm oscillations of four-probe resistance in topological quantum rings in silicene and bilayer graphene}

\author{Bart\l{}omiej Rzeszotarski}

\affiliation{AGH University of Science and Technology, Faculty of Physics and
Applied Computer Science,\\
 al. Mickiewicza 30, 30-059 Kraków, Poland}

\author{Alina Mre\'n{}ca-Kolasi\'n{}ska}

\affiliation{AGH University of Science and Technology, Faculty of Physics and
Applied Computer Science,\\
 al. Mickiewicza 30, 30-059 Kraków, Poland}

\author{Bart\l{}omiej Szafran}

\affiliation{AGH University of Science and Technology, Faculty of Physics and
Applied Computer Science,\\
 al. Mickiewicza 30, 30-059 Kraków, Poland}

\begin{abstract}
We consider observation of Aharonov-Bohm oscillations in clean systems 
based on the flow of topologically protected currents in silicene and bilayer graphene.
 The chiral channels in these materials are defined by the flips of the vertical electric field. 
The line of the flip confines chiral currents flowing along it in the direction determined by the valley.
 We present an electric field profile that forms a crossed ring to which four terminals can be attached, and find that the conductance matrix elements oscillate in the perpendicular magnetic field in spite of the absence of backscattering. We propose a four-probe resistance measurement setup, and demonstrate that the resistance oscillations have large visibility provided that the system is prepared in such a way that a direct transfer of the chiral carriers between 
the current probes is forbidden. 
\end{abstract}
\maketitle
\section{Introduction}

In III-V systems with the two-dimensional electron gas the Aharonov-Bohm interferometers
are formed by definition of gated channels forming ring-like structures by etching \cite{elia}  or surface oxidation \cite{so}.
Quantum rings are also defined in graphene by etching \cite{grae}. In the etched systems disorder and resulting electron backscattering within the 
arms of the ring are usually present which lowers the visibility of the Aharonov-Bohm conductance oscillations \cite{Szapo}. In the present work we consider Aharonov-Bohm
interferometers with arms formed by chiral \cite{chiral} channels that are protected against backscattering by symmetry constraints. 

Graphene \cite{graphene} nanoribbon  \cite{nrbr} with  zigzag edges forms a perfect chiral channel
for low Fermi energy.  The Fermi wave vectors corresponding to the current flow in one direction or the other appear in opposite valley states \cite{zz1,zz2,zz3}. The chiral \cite{chiral} valley current within a quasi one-dimensional channel is protected against backscattering by a smooth potential variation. Only potential defects that are short range on the scale of the lattice constant
 can induce intervalley transition that implies  backscattering  \cite{zz1,zz2,zz3}. 
However, formation of a quantum ring of purely zigzag edges is unlikely. For that reason we consider
 chiral channels  defined within the bulk of the sample by gating. 
In staggered monolayer graphene \cite{niu,prl18},
 in buckled silicene lattice \cite{Aufray10,Liu11,Liu,chow},
or other 2D Xene materials \cite{xene,germanen,stanene}, 
the chiral channels for the electron flow can be tailored  by a symmetry breaking potential along its zero lines \cite{niu,Ezawa2012a,Szafran19}. 
For buckled silicene \cite{Aufray10,Liu11,Liu,chow} -- a hexagonal crystal with the two sublattices placed on two parallel planes --  the symmetry breaking potential is introduced by perpendicular electric field
\cite{Ezawa2012a,Szafran19}.  Similar chiral channels appear in bilayer graphene \cite{morpugo,macdo,peet,down,muktu2,prl18} along the flip of the vertical electric field or
in bilayer graphene  at  the $\mathtt{AB}$/$\mathtt{BA}$ stacking interface induced by 
a dislocation  \cite{down,prx}  or twist of the layers \cite{twi,margi}.
The $\mathtt{AB}$/$\mathtt{BA}$ interfaces in twisted bilayer graphene
form a triangular lattice with the direction of the current flow opposite for both valleys \cite{margi,macdo2,bigf1}. 

Recently, a ring-like system with splittings of the chiral zero-line channels was proposed for both silicene and bilayer graphene \cite{Szafran19}. The
electron passage time across this system is a periodic function of the external magnetic field 
due to the Aharonov-Bohm phase difference accumulated from the vector potential \cite{Szafran19}. 
However, the two-terminal Landauer conductance of these systems is independent of the external magnetic field, since the backscattering is absent due to the valley protection. 
In order to observe the Aharonov-Bohm oscillations in two-terminal conductance one should rely on
atomic-scale disorder. The atomic disorder is hard to  control and electron interferometers should be difficult to construct in this way.

The message of this paper is that
one can design a four-terminal interferometer device 
for the observation of the Aharonov-Bohm oscillations of conductance for clean chiral
channels defined in both silicene
and bilayer graphene that works in absence of the electron backscattering. 
We study the chiral current flow in the channels formed by the electric potential flips that define the four-terminal crossed-quantum ring in silicene and bilayer graphene. A simulation of two nonequivalent four-point resistance  measurement setups is perfomed. 
 We find a distinct Aharonov-Bohm periodicity in the resistance amplitude that is associated with the interference on 1/4 of the ring area.
We discuss the interference paths that are behind this periodicity.

\section{Theory}
\subsection{Silicene}\label{sec:theor_silic}
In this work we use the atomistic tight-binding Hamiltionian spanned by $p_z$ orbitals \cite{Liu,Ezawa,chow} 
\begin{align}
H=&-t\sum_{\langle n,m\rangle }\left(  \mathrm{p}_{nm} c_{n}^\dagger c_{m} +h.c.\right)
+ \sum_{n} V(\mathbf{r}_n)c^\dagger_{n}c_{n}  \label{eq:HH}
\end{align}
where $\langle n,m\rangle $ stands for the nearest neighbor ions. The $c_{n}^\dagger$ ($c_{n}$) is the creation (annihilation) operator for an electron on site $n$, and $t=1.6$ eV is the hopping parameter \cite{Liu,Ezawa}.
We introduce the magnetic field via  the Peierls phase in the $\mathrm{p}_{nm}$ term, where 
$\mathrm{p}_{nm}=e^{i\frac{e}{\hbar}\int_{\vec{r_n}}^{\vec{r_m}}\vec {\mathbf{A}}\cdot \vec {dl}}$ with the vector potential $\vec{\mathbf{A}} = ( 0,\mathcal{B}x,0)$.
The crystal lattice vectors ${\bf a}_1=a \left(\frac{1}{2},\frac{\sqrt{3}}{2},0\right)$
and ${\bf a}_2=a \left(1,0,0\right)$ define the positions ${\bf r}_{\bf n}^A=n_1 {\bf a}_1+n_2 {\bf a}_2$   of the ions of the $\mathtt{A}$ subblatice, where the silicene lattice constant $a=3.89$ \AA, and $n_1$, $n_2$ are integers, and the ions of the $\mathtt{B}$ sublattice are shifted by the basis vector $(0,d,\tau)$, where $d=2.248$
 \AA\;
is the nearest neighbor in-plane distance and $\tau=0.46$ \AA\; is the vertical shift between the sublattice planes.

\begin{figure}[htbp]
\centering
\includegraphics[width=0.24\textwidth,trim=0 -400 0 0 ,clip]{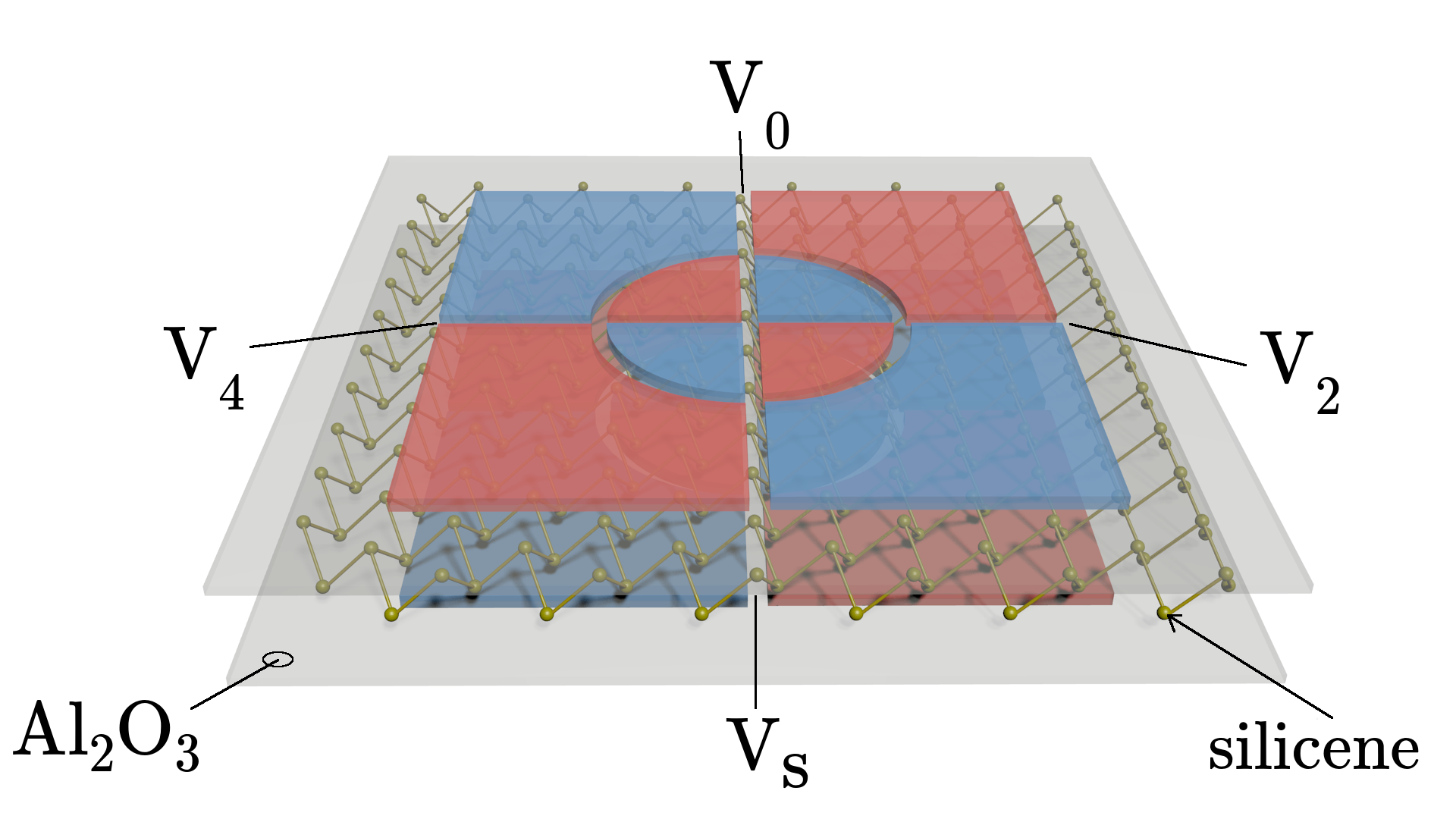}\includegraphics[width=0.24\textwidth]{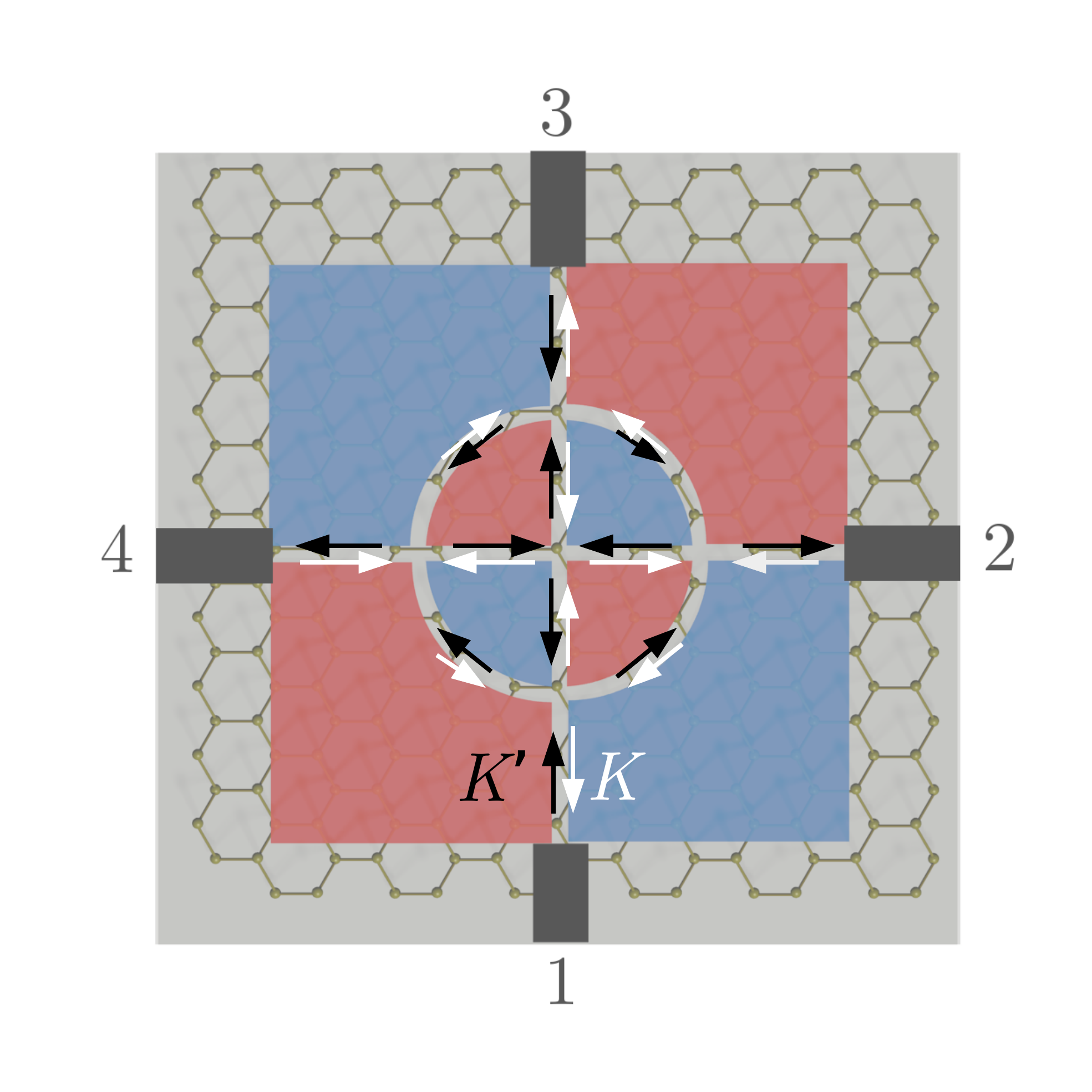}
\put(-245,100){(a)}\put(-125,100){(b)}
\caption{ (a) Sketch of the silicene monolayer between top and bottom gates. Positive potential $V_G$ is put on the blue gates, and negative $-V_G$ on the red gates. (b) Top view of the system with leads numeration. Black (white) arrows denote for the directions of $K'$ ($K$) valley protected charge currents within each channel defined by gate interfaces.  }
\label{fig:sch1}
\end{figure}

The quantum ring with the chiral channels is formed by the electric field induced by the split top and bottom gates [Fig. \ref{fig:sch1}(a)].
The systems of multiple dual gates below and above the two dimensional crystals 
are used to modify the local electron structure \cite{ga1,ga2,ga3,bigf1}. 
 The inversion of the field creates a topologically protected conducting channel. We consider a ring of radius $\mathcal{R}$ with the center at the origin formed by the model potential
\begin{equation}
V_{\mathtt{A}} = \frac{8V_G}{\pi^3} \arctan{\left(\frac{x}{\lambda}\right)} \arctan{\left(\frac{y}{\lambda}\right)} \arctan{\left(\frac{\mathcal{R}-r}{\lambda}\right)}, \label{eq:VA}
\end{equation}
where $r=\sqrt{x^2+y^2}$ is the distance from the origin, $V_G$ is the gate potential and $\lambda$ is the parameter responsible for the inversion length.
For symmetric gating the potential on the  $\mathtt{B}$ sublattice is opposite $V_{\mathtt{B}}(\mathbf{r}) = - V_{\mathtt{A}}(\mathbf{r})$.

\begin{figure}
  \includegraphics[width=0.5\textwidth]{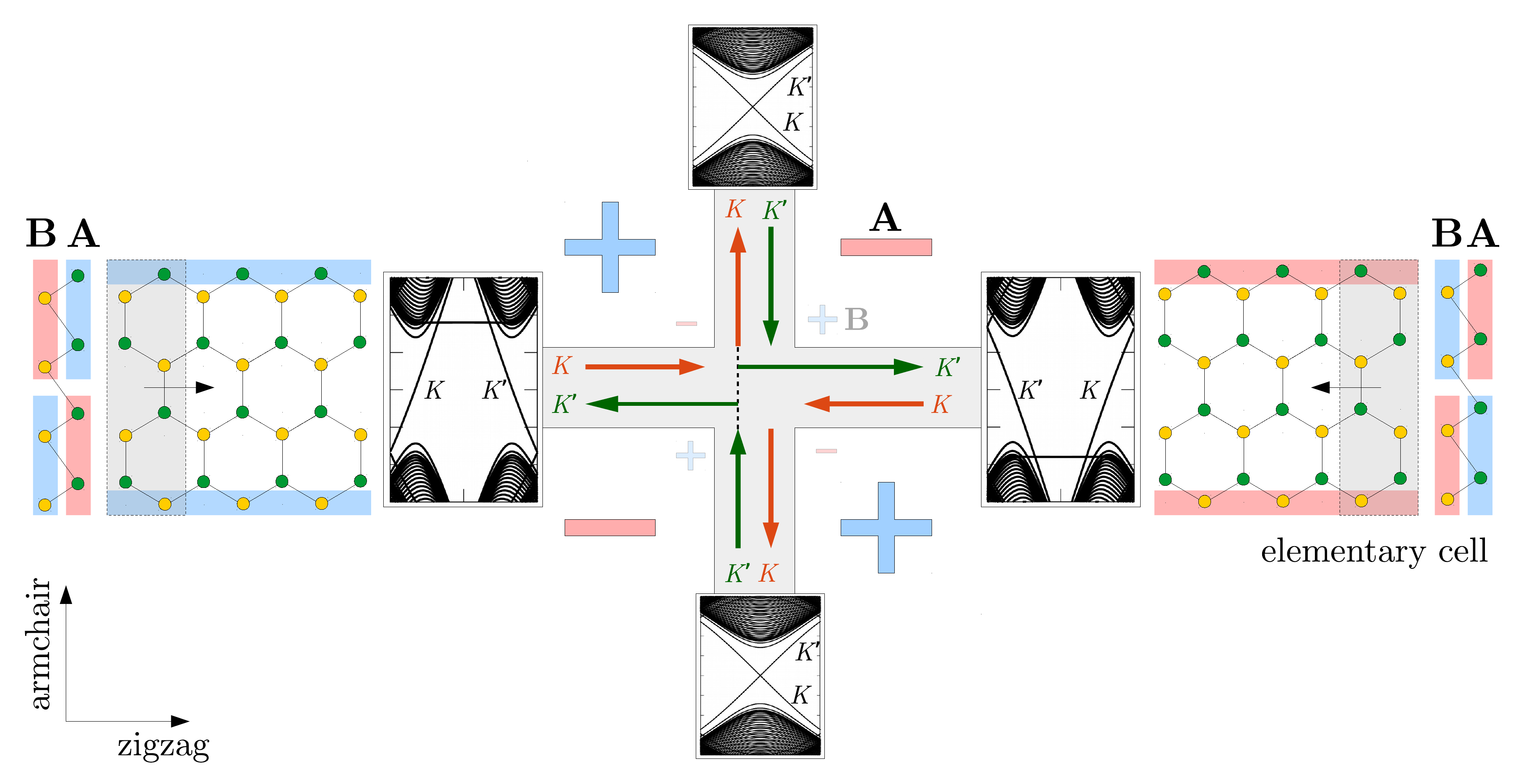}
  \caption{Schematic view of a 4-terminal crossing channels defined by flips of the electric potential. Negative potential -$V_G$ is shown by red color ("$-$" symbol) and positive $V_G$ by blue color ("$+$" symbol), for both the $\mathtt{A}$ (upper) and $\mathtt{B}$ (lower) sublattice. Green (orange) arrows indicate the orientation of the $K'$ ($K$) currents in the channel that are associated with a specific valley marked in the band structures for each lead. In the zigzag ribbon the potential at the edge changes its sign and shifts the flat subbands -- corresponding to the edge states -- to different Fermi levels, $E_F = +V_G$ and $E_F = -V_G$, respectively.}
  \label{fig:sch1_vb}
\end{figure}

In Fig. \ref{fig:sch1}(b) we plot the direction of the current channels that are 
open for the $K$ and $K'$ valley electron flow. The $K'$ valley electrons move along
the zero line of the potential given by Eq.~(\ref{eq:VA}) leaving the region
of negative potential on the $\mathtt{A}$ sublattice on the left hand-side.
Note that the current injected from terminal zero-line 1 can be directed
to either terminal 2 or terminal 4. In terminal 3 there is no $K'$ valley state that carries the electron flow up, away from the ring (Fig. \ref{fig:sch1_vb}). 
In every channel the direction of the current flow for the $K$ valley (with respect to the $K'$) is opposite.

\begin{figure}[htbp]
\centering
\includegraphics[width=0.21\textwidth,trim=0 0 0 0 ,clip]{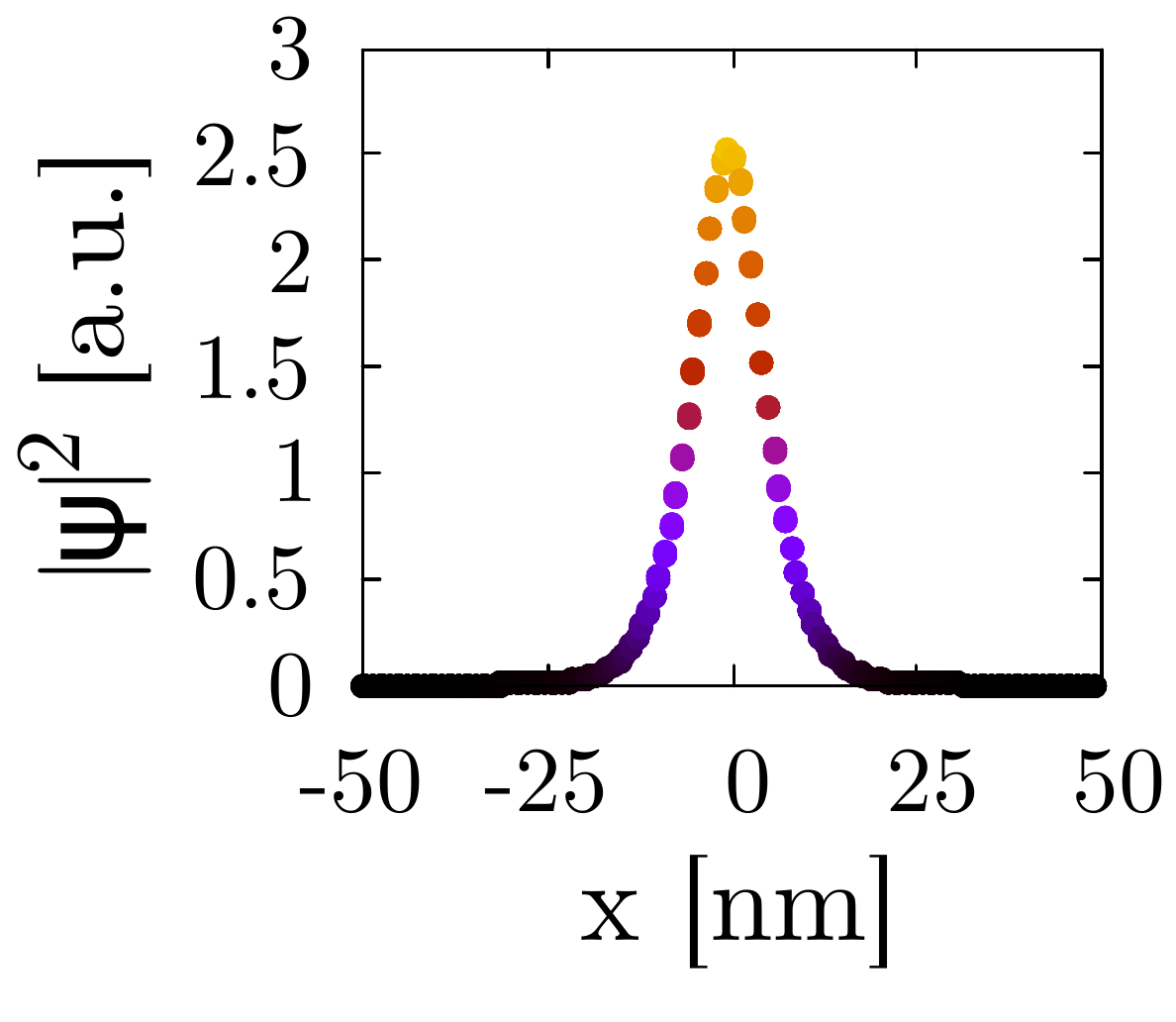}\includegraphics[width=0.2\textwidth]{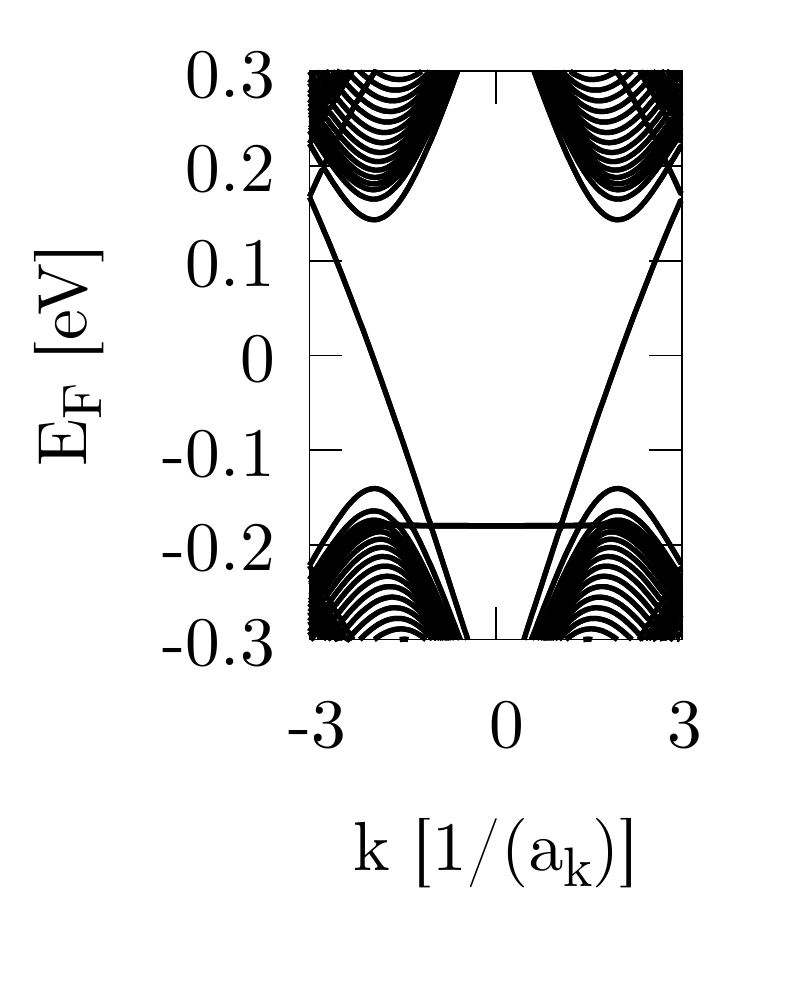}
\put(-200,100){(a)}\put(-100,100){(b)}\put(-53,80){$K'$}\put(-30,80){$K$}
\caption{ (a) Probability density $|\Psi |^2$ of the states confined along the zero lines and (b) band structure for the input lead 1 with zigzag edges and electric potential defined as described in Fig. \ref{fig:sch1}.  
The flat horizontal band corresponds to the edge state. The translation vector of the supercell of the zigzag nanoribbon has the length of two lattice constants $a_k = 2a$.}
\label{fig:bnd}
\end{figure}

The wave function of the states confined laterally at the zero line 
near $x=0$ is plotted in Fig. \ref{fig:bnd}(a).  
The confined states correspond to linear bands that appear within the energy gap  [Fig. \ref{fig:bnd}(b)].

\subsection{Bilayer graphene}

For bilayer graphene the inversion-symmetry-breaking potential can be introduced by an electric field perpendicular to the sheet. We consider a bilayer-graphene-based system analogous to the one described in Sec.~\ref{sec:theor_silic}, with the difference that due to the presence of two layers, two topological states occur instead of one as in silicene. 
We consider the tight-binding Hamiltonian  similar as in Eq.~(\ref{eq:HH})
\begin{equation}
   H=-\sum_{\langle n,m\rangle } \left( t_{nm} \mathrm{p}_{nm} c_n^\dagger c_m+h.c.\right)+\sum_n V({\bf r}_n) c_n^\dagger c_n, 
\label{eq:BLG_HH}
\end{equation}
with graphene lattice constant $a_{CC}=1.42$ {}\AA, the interlayer distance of $d=3.35${}\AA\; and the tight-binding parameters of bilayer graphene with Bernal stacking \cite{Partoens}, where 
$t_{nm}=-3.12$ eV for the nearest neighbors within the same layer, 
and for the interlayer coupling, $t_{nm}=-0.377$ eV for the $\mathtt{A-B}$ dimers, $t_{nm}=-0.29$ eV for the skew hoppings between atoms of the same sublattice, and $t_{nm}=0.12$ eV -- between the atoms of different sublattices. 

The model potential in the lower layer is described by a formula analogous as in Eq.~(\ref{eq:VA}), and in the upper layer it has the opposite polarization, but the sign is the same in both sublattices within the same layer. We use $\lambda=4$ nm, $\mathcal{R}=100$ nm, and $V_G=200$ meV.

\subsection{Landauer approach}
We solve the electron scattering problem formed in the tight binding model with the wave-function matching (WFM) technique. The details of the method were described in Refs. \cite{bubel,rzeszotarski18}. The electron transfer probability is calculated as
\begin{equation}
T^w_{\xi\eta} = \sum_v |t^{wv}_{\xi\eta}|^2,
\end{equation}
where $t^{wv}_{\xi\eta}$ denote  the probability amplitude for the transfer from incoming mode $v$ in the input lead $\eta$ to  outgoing mode $w$ in the output lead $\xi$. Thus, the Landauer conductance formula for the transfer from lead $\eta$ to $\xi$ can be written as
\begin{equation}
G_{\xi\eta} = G_0 \sum_{w} T^w_{\xi\eta},
\end{equation} 
where $G_0 = e^2/h$ is the flux quantum. 

We focus our attention on the Fermi level $E_F \in \left\{ 0:0.1\right\}$ eV and take into account the spin degree of freedom so that all the assumptions provide $\mathcal{G}=\max(G_{\xi\eta})=2G_0$ for silicene, and $\mathcal{G}=\max(G_{\xi\eta})=4G_0$ for bilayer graphene.




\subsection{Conductance matrix }
The scattering problem for the four-terminal system was solved for each lead as an input channel and the results were collected in the conductance matrix $\mathbf{G}$ with the general form
\begin{equation}
\mathbf{G} = \begin{pmatrix}
g_{11} & -G_{12} & -G_{13} & -G_{14} \\
-G_{21} & g_{22} & -G_{23} & -G_{24} \\
-G_{31} & -G_{32} & g_{33} & -G_{34} \\
-G_{41} & -G_{42} & -G_{43} & g_{44}
\end{pmatrix},
\label{eq:cm}
\end{equation}
with $g_{ii}=\sum_{j\neq i}G_{ij}$.
Due to the rotational symmetry ($C_4$ in terms of channel shape) the conductance matrix $\mathbf{G}$ can be put in the form
\begin{equation}
\mathbf{G} = \begin{pmatrix}
\mathcal{G} & -B 			& 0 		& -A \\
-A 			& \mathcal{G} 	& -B 		& 0 \\
0 			& -A 			& \mathcal{G} & -B \\
-B 			& 0 			& -A 		& \mathcal{G}
\end{pmatrix},
\label{eq:simply}
\end{equation}
where the coefficients 
\begin{align}
A = G_{14} = G_{21} = G_{32} = G_{43} \\
B = G_{41} = G_{12} = G_{23} = G_{34}\\
0 = G_{13} = G_{24} = G_{31} = G_{42}
\end{align}
and $B = \mathcal{G}-A$.


Assuming that $V_3=0$ we can truncate the 3\textit{rd} column \cite{Datta1995} and calculate the resistance matrix $\mathbf{R} = \mathbf{G}^{-1}$ that can be written as follows

\begin{align}
\mathbf{R} = &\begin{pmatrix}
R_{11} & R_{12} & R_{14}  \\
R_{21} & R_{22} & R_{24}  \\
R_{41} & R_{42} & R_{44}
\end{pmatrix}
 = \frac{1}{W}\begin{pmatrix}
\mathcal{G}^2 	& B \mathcal{G}				& A \mathcal{G}  \\
A  \mathcal{G}				& \mathcal{G}^2-AB 	& A^2  \\
B \mathcal{G}				& B^2 				& \mathcal{G}^2-AB
\end{pmatrix}
\label{eq:rmat}
\end{align}
with matrix determinant $W = \mathcal{G}(B^2+A^2)$ that is always positive.

\subsection{4-point resistance measurement}

\begin{figure}[htbp]
\centering
\includegraphics[width=0.24\textwidth]{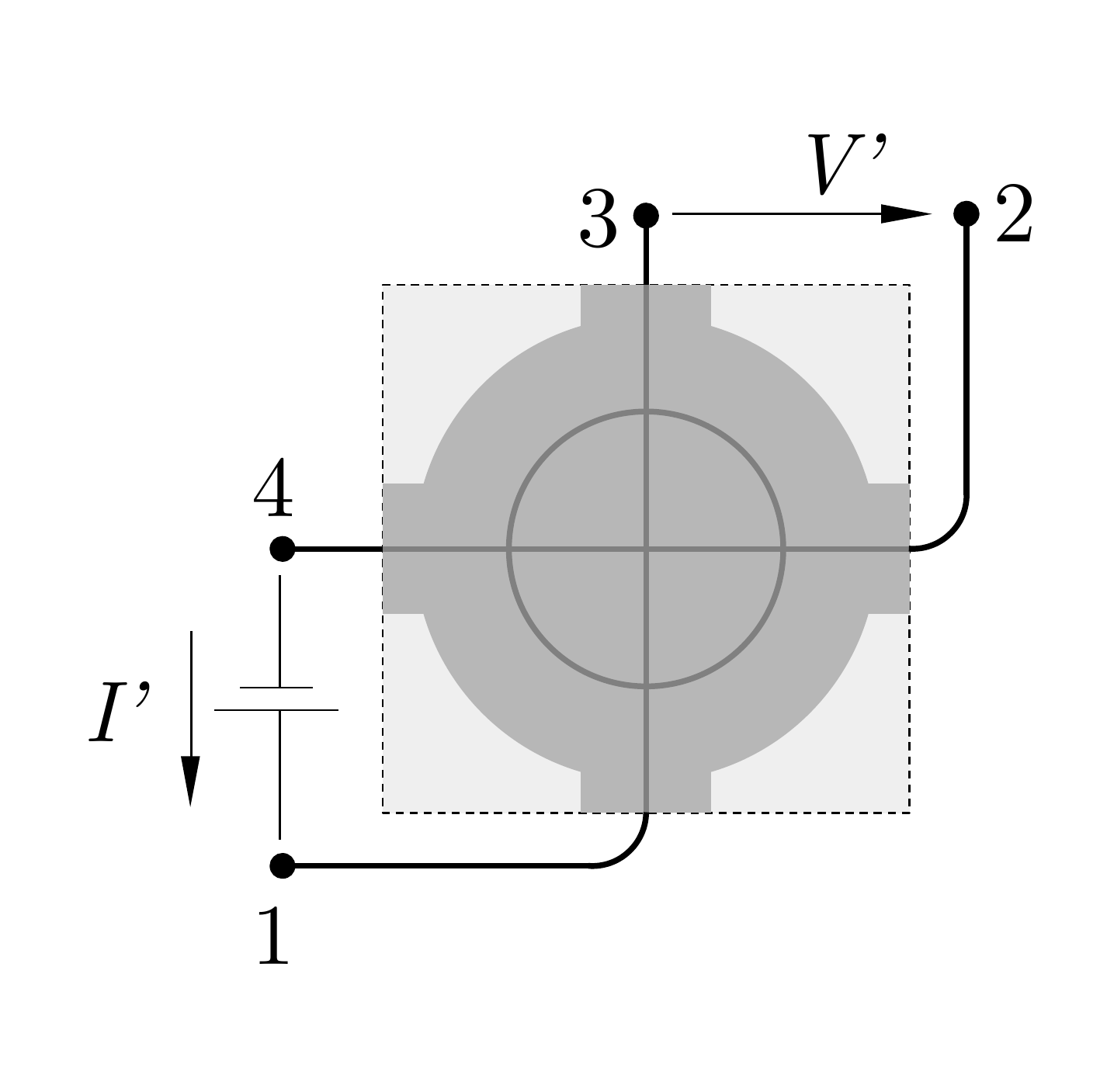}\includegraphics[width=0.24\textwidth]{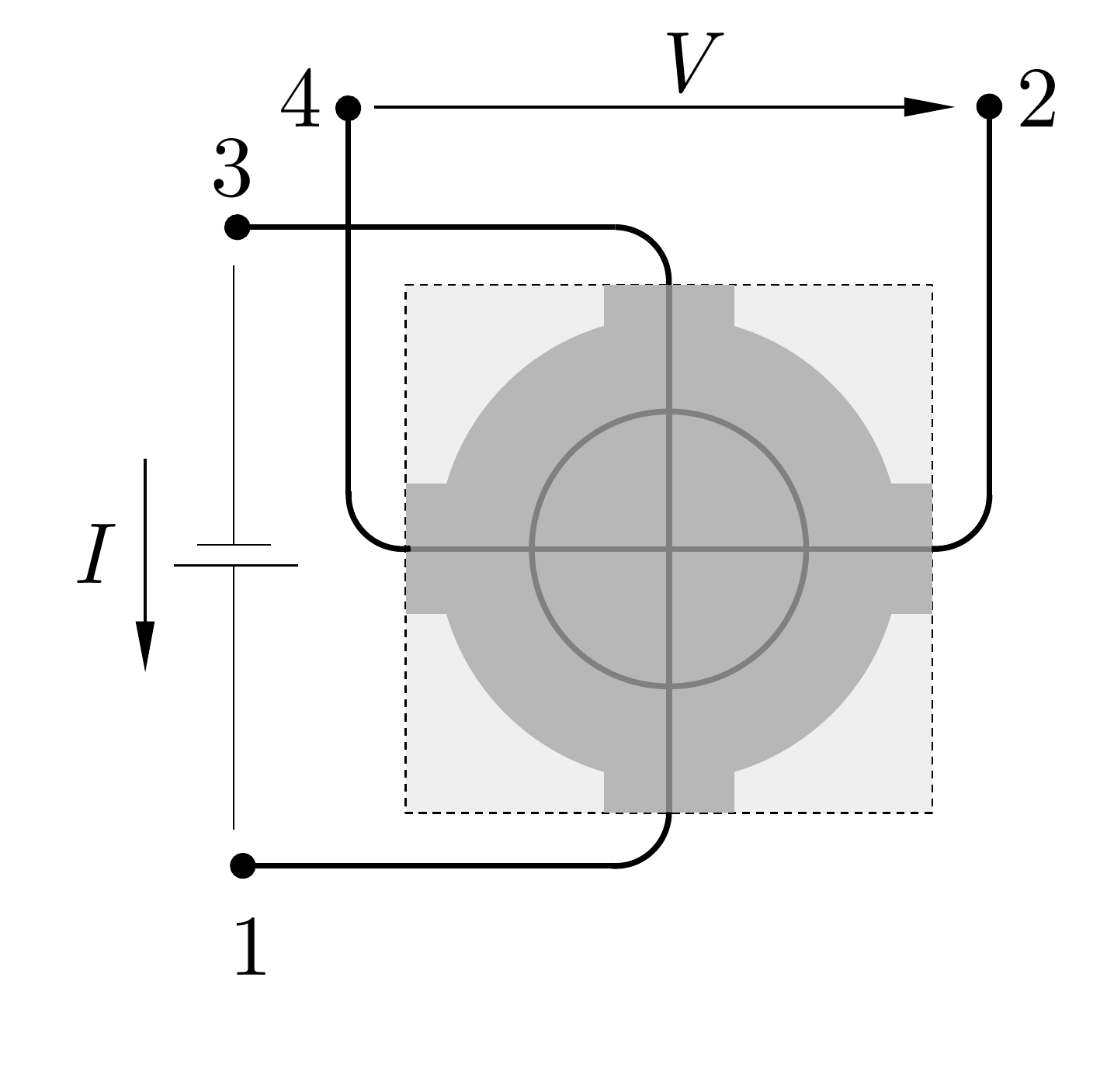}
\put(-245,110){(a)}\put(-125,110){(b)}
\caption{ Schemes for experimental resistance measurement configuration: for the current flow between neighbour (a)  and  opposite (b) leads. In both configurations the voltage $V_3$ associated to the $3$rd lead is set to 0. }
\label{fig:conf}
\end{figure}

We consider two configurations of resistance measurement (Fig. \ref{fig:conf}) in the system with varied voltage and current terminals.

For the first configuration from Fig. \ref{fig:conf}(a) the resistance 
is calculated as 
\begin{equation}
R' = \frac{V'}{I'}=\left[ \frac{V_2}{I_1} \right]_{\substack{I_2=0\\I_1=-I_4}} = R_{21} - R_{24} = \frac{AB}{W}
\label{eq:R'}
\end{equation}
and for the other [Fig. \ref{fig:conf}(b)],

\begin{equation}
R = \frac{V}{I}=\left[ \frac{V_4-V_2}{I_1} \right]_{\substack{I_2=I_4=0}} = R_{41} - R_{21} = \frac{\mathcal{G}\left( B-A \right)}{W}
\label{eq:R}
\end{equation}

\section{Results and Discussion}
\subsection{Small ring $\mathcal{R}=100\ \mathrm{nm}$}
\begin{figure}[htbp]
\centering

\includegraphics[width=0.48\textwidth]{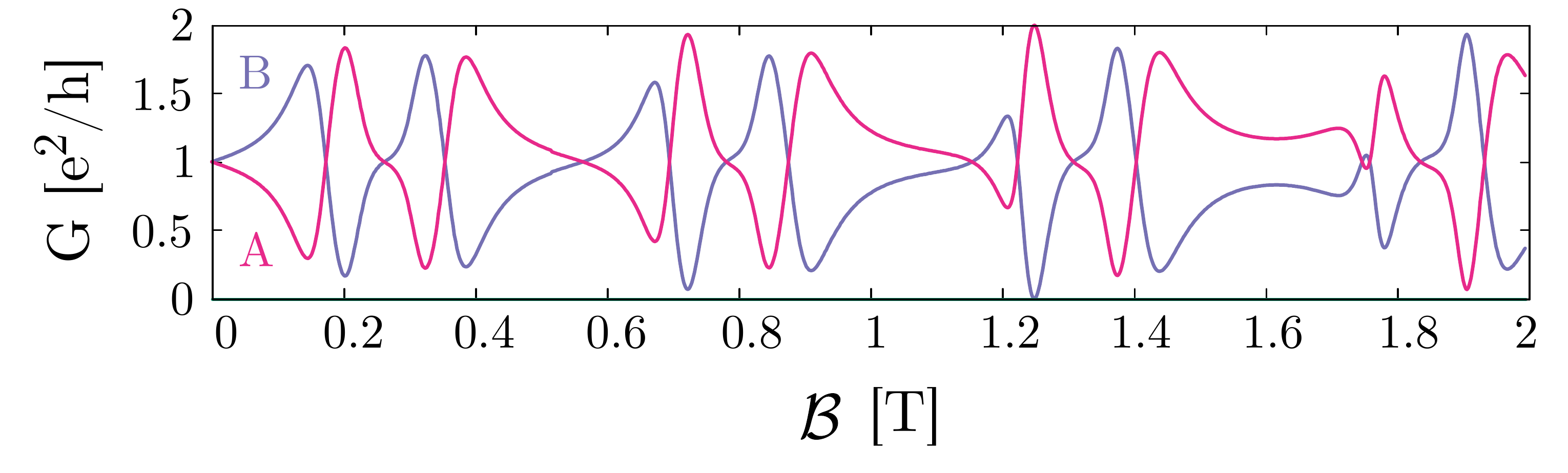}\put(-240,75){(a)}

\includegraphics[width=0.48\textwidth,trim=0 0 0 0 ]{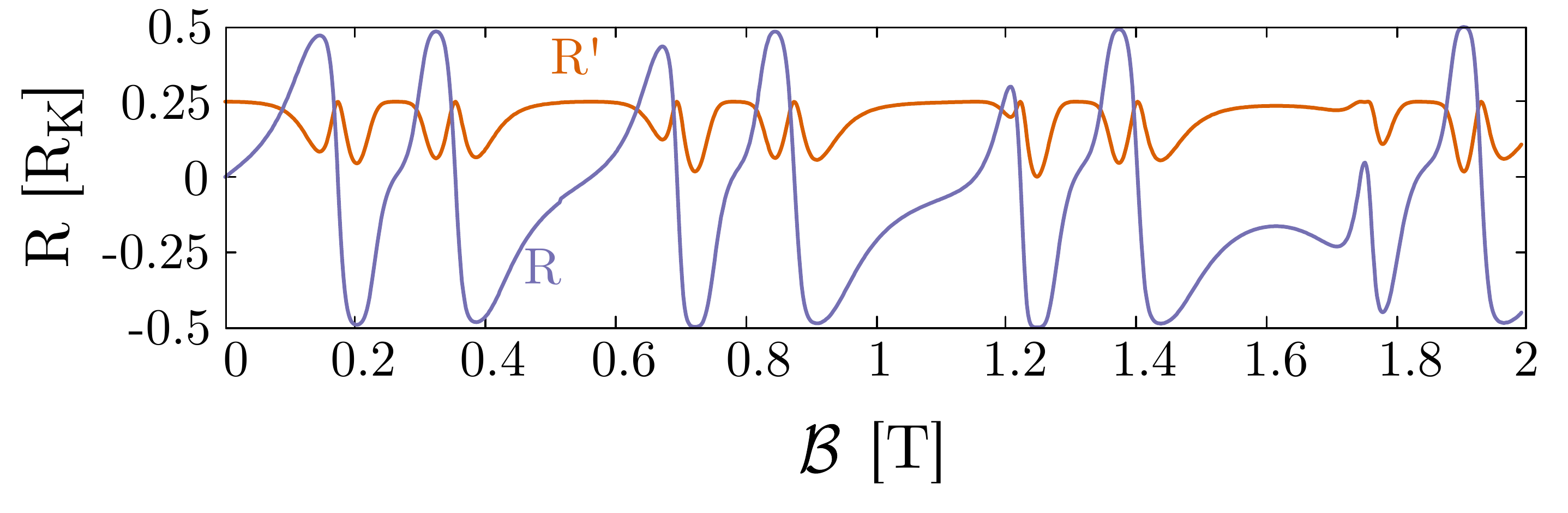}\put(-240,75){(b)}

\caption{  (a) Conductance plot for simplified conductance matrix elements (Eq.~\ref{eq:simply}) for the system in external magnetic field at fixed Fermi level $E_F$ = 20 meV. (b) Resistance $R'$ for the case in Fig.~\ref{fig:conf}(a) and $R$ [case Fig.~\ref{fig:conf}(b)] in units of von Klitzing constant $R_K$. }
\label{fig:gab_2}
\end{figure}

\begin{figure}[htbp]
\centering

\includegraphics[width=0.48\textwidth]{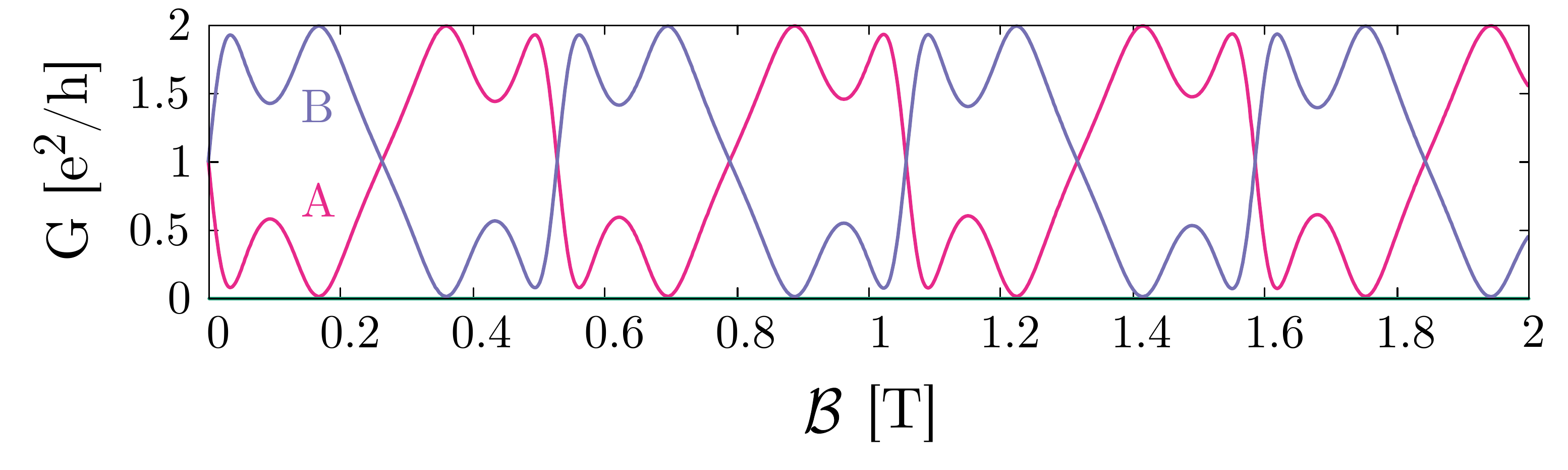}\put(-240,75){(a)}

\includegraphics[width=0.48\textwidth,trim=0 0 0 0 ]{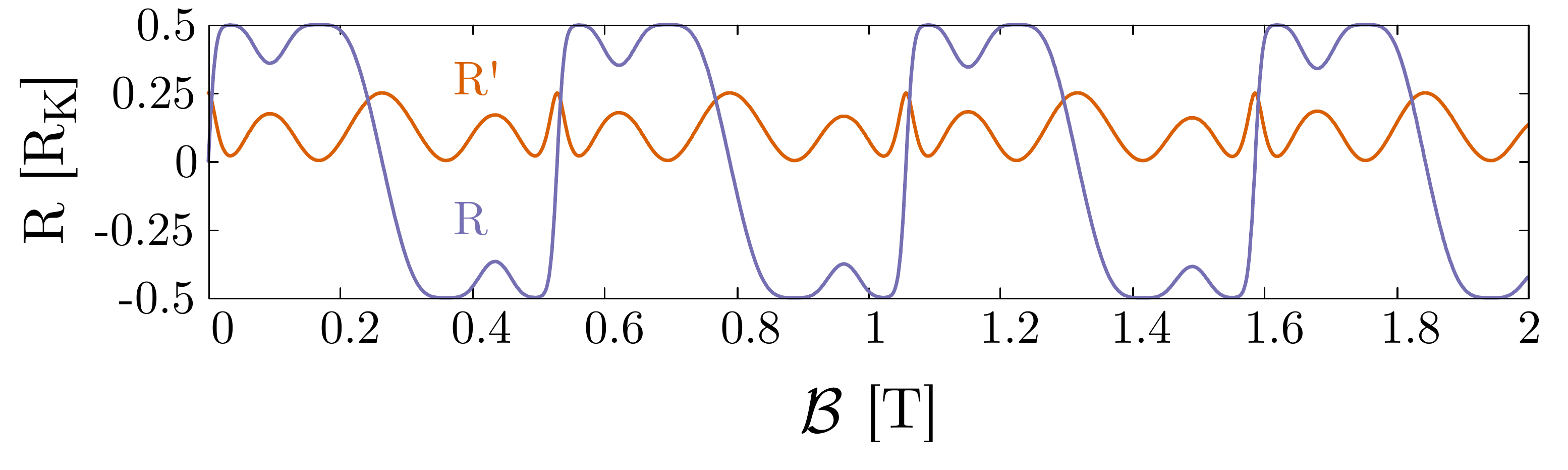}\put(-240,75){(b)}

\caption{ (a) Conductance and (b) resistance plots same as Fig. \ref{fig:gab_2} but for $E_F$ = 6.43 meV.  }
\label{fig:gab_642}
\end{figure}

In this subsection for the silicene system we use $V_G=200$ meV and $\lambda=4$ nm.
In Fig. \ref{fig:gab_2} and \ref{fig:gab_642} we plotted
the results for the conductance matrix elements (upper plots) 
and the resistances $R$ and $R'$ (lower plots) for $E_F=20$ meV and $E_F=6.43$ meV, respectively. 
The oscillations of the matrix elements that have nearly maximal amplitude
are translated to oscillations of resistance that have high  ($R$) 
or low ($R'$) visibility.

The current probe terminals 
for configuration $R'$ correspond to an open direct current path.
The $R'$ resistance has a constant sign since the numerator in 
Eq. (\ref{eq:R'}) is always nonnegative $AB\geq 0$.

For configuration $R$ the current from terminal 1 can reach the terminal 3  only via the voltage terminals 2 and 4 which absorb the current and send an equal current back in the opposite valley, which is necessary to keep the net current at the voltage  probes equal to zero. The resistance $R$ changes sign [Fig. \ref{fig:gab_2}(b) and Fig. \ref{fig:gab_642}(b)] as the magnetic field is varied. 
From Eq.~(\ref{eq:R}), since the determinant $W$ is positive,  the sign change needs to be accompanied by the sign change of the difference $V_4-V_2$ (or the matrix elements $B-A=G_{41}-G_{14}$). Hence, the changes of sign of $R$ appear when the electron transfer probability from terminal 1 to 4 
crosses the electron transfer probability in the opposite direction. 
The directions become non-equivalent from the point of view of the electron transfer when the external magnetic field is introduced. 

\begin{figure}[htbp]
\centering

\includegraphics[scale=0.25,trim=80 52 100 0 ,clip]{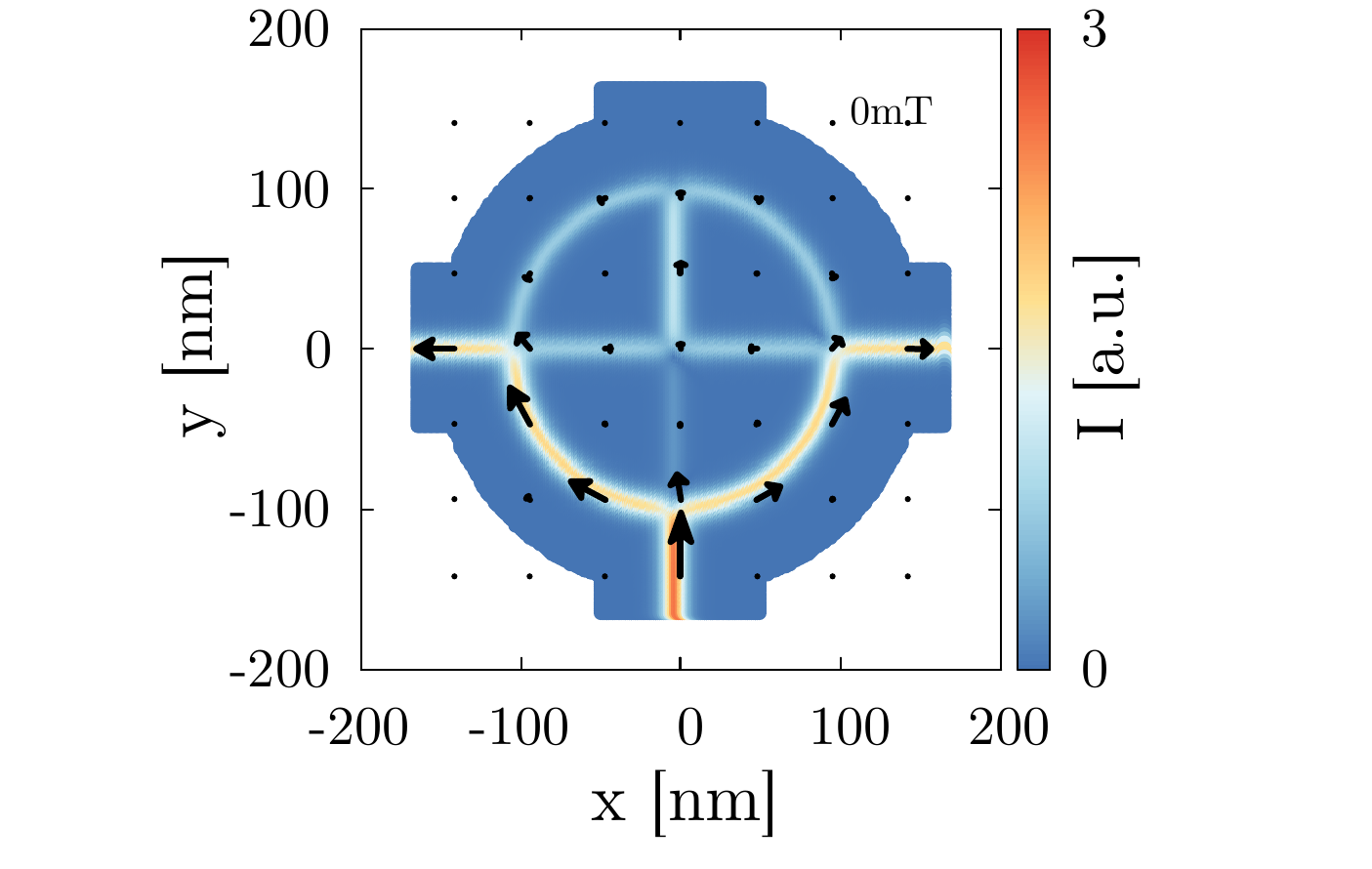}\put(-100,80){(a)}
\includegraphics[scale=0.25,trim=115 52 100 0 ,clip]{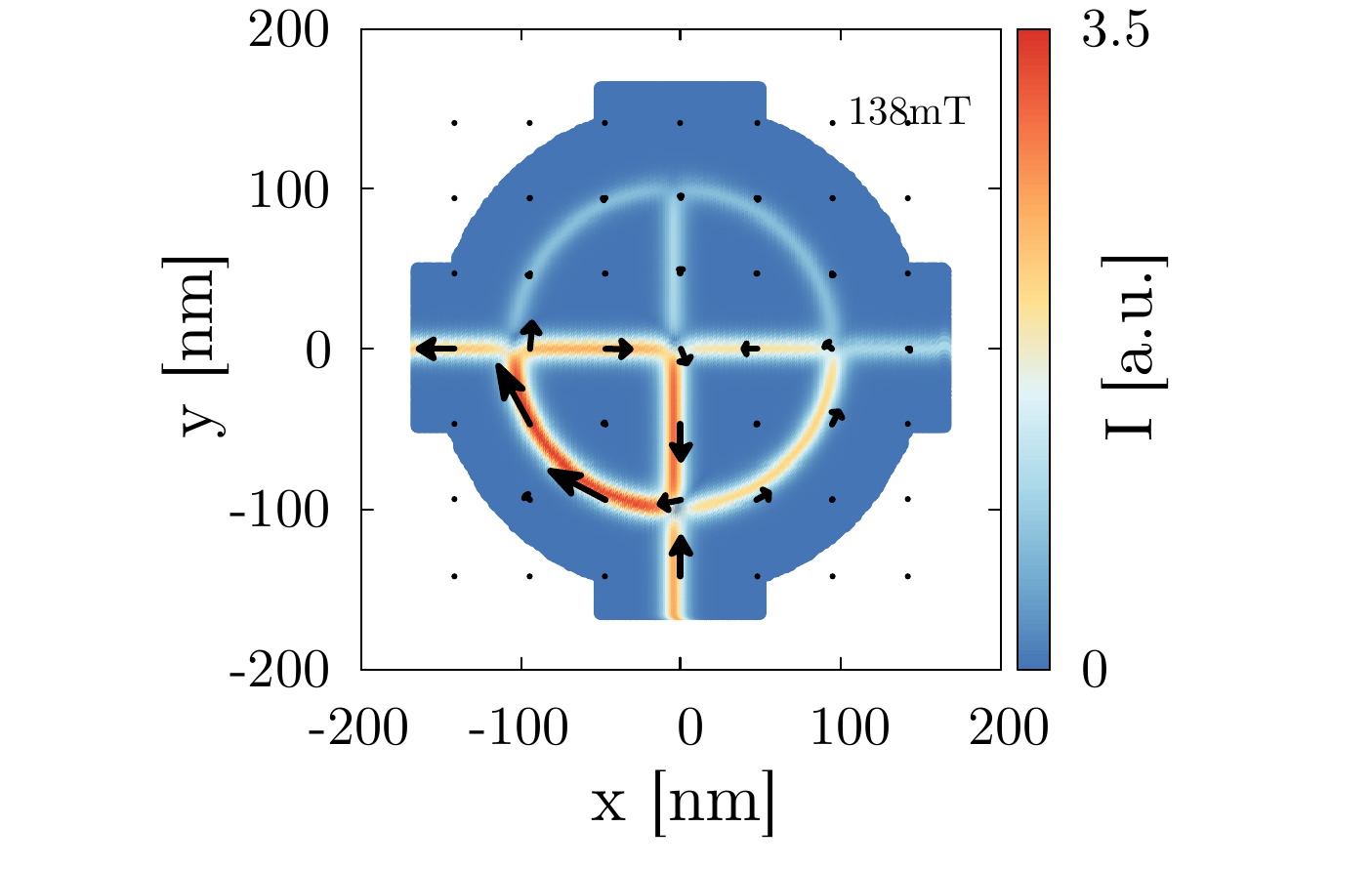}\put(-100,80){(b)}

\includegraphics[scale=0.25,trim=80 52 100 0 ,clip]{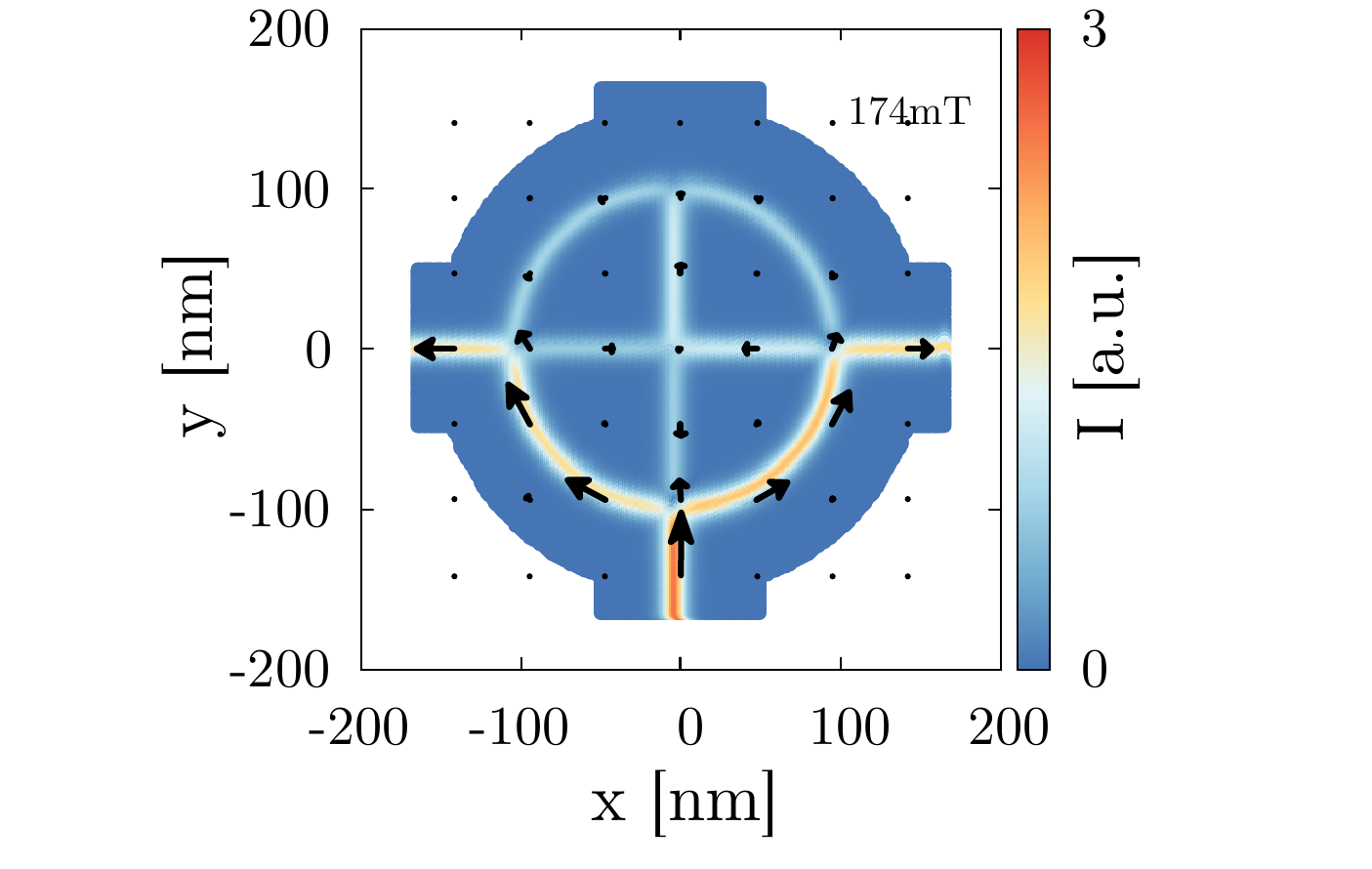}\put(-100,80){(c)}
\includegraphics[scale=0.25,trim=115 52 100 0 ,clip]{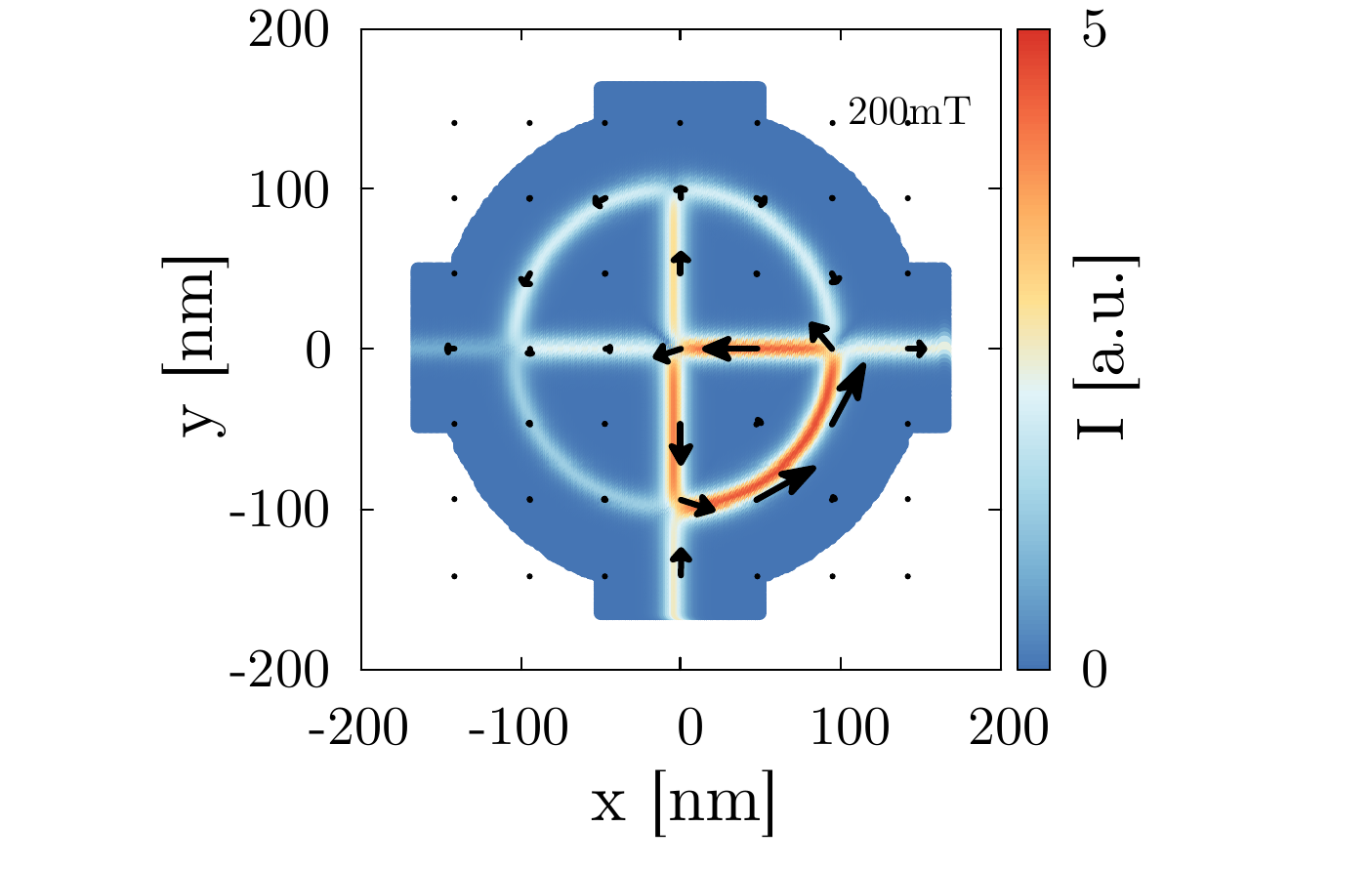}\put(-100,80){(d)}

\includegraphics[scale=0.25,trim=80 0 100 0 ,clip]{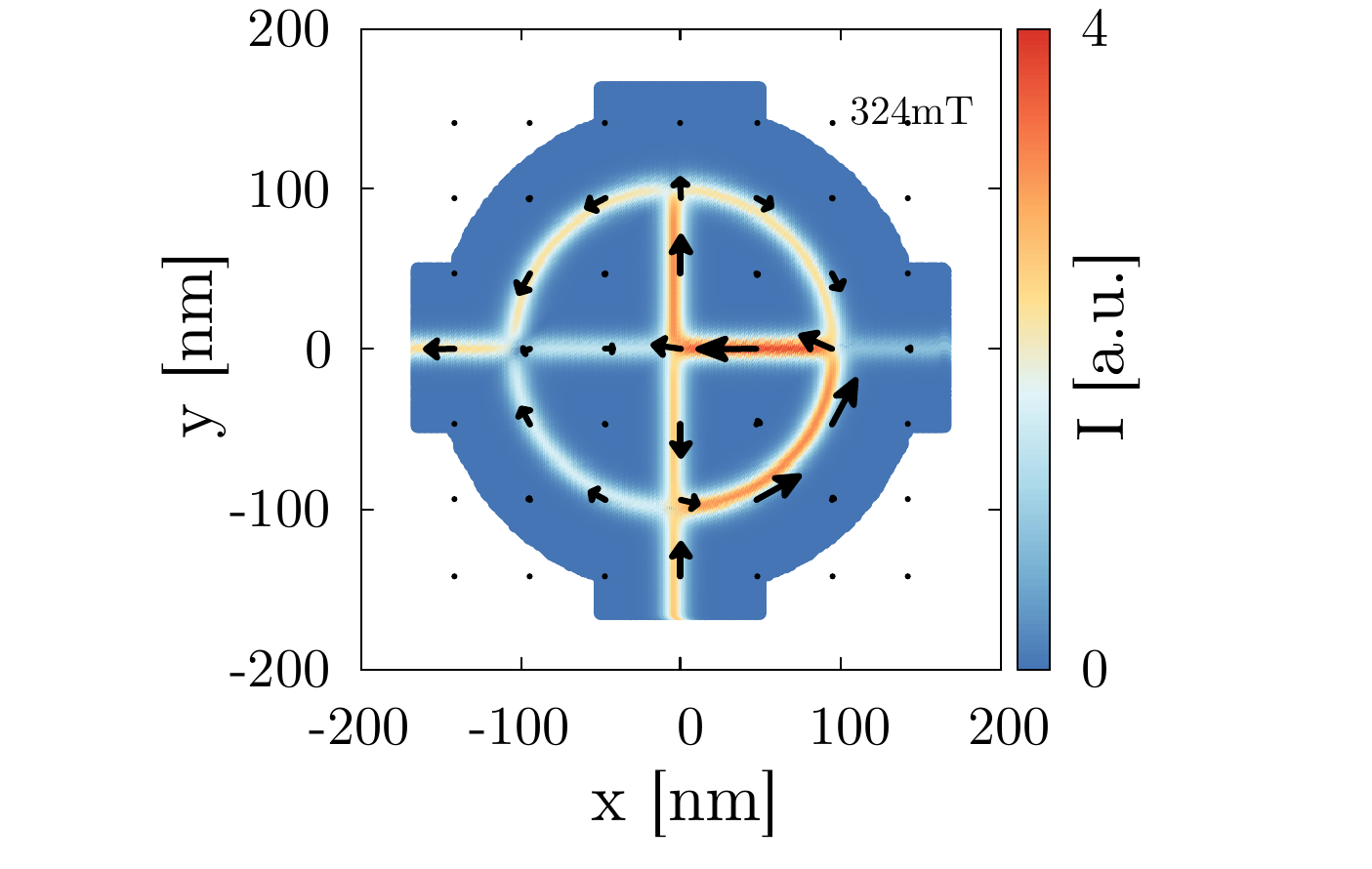}\put(-100,95){(e)}
\includegraphics[scale=0.25,trim=115 0 100 0 ,clip]{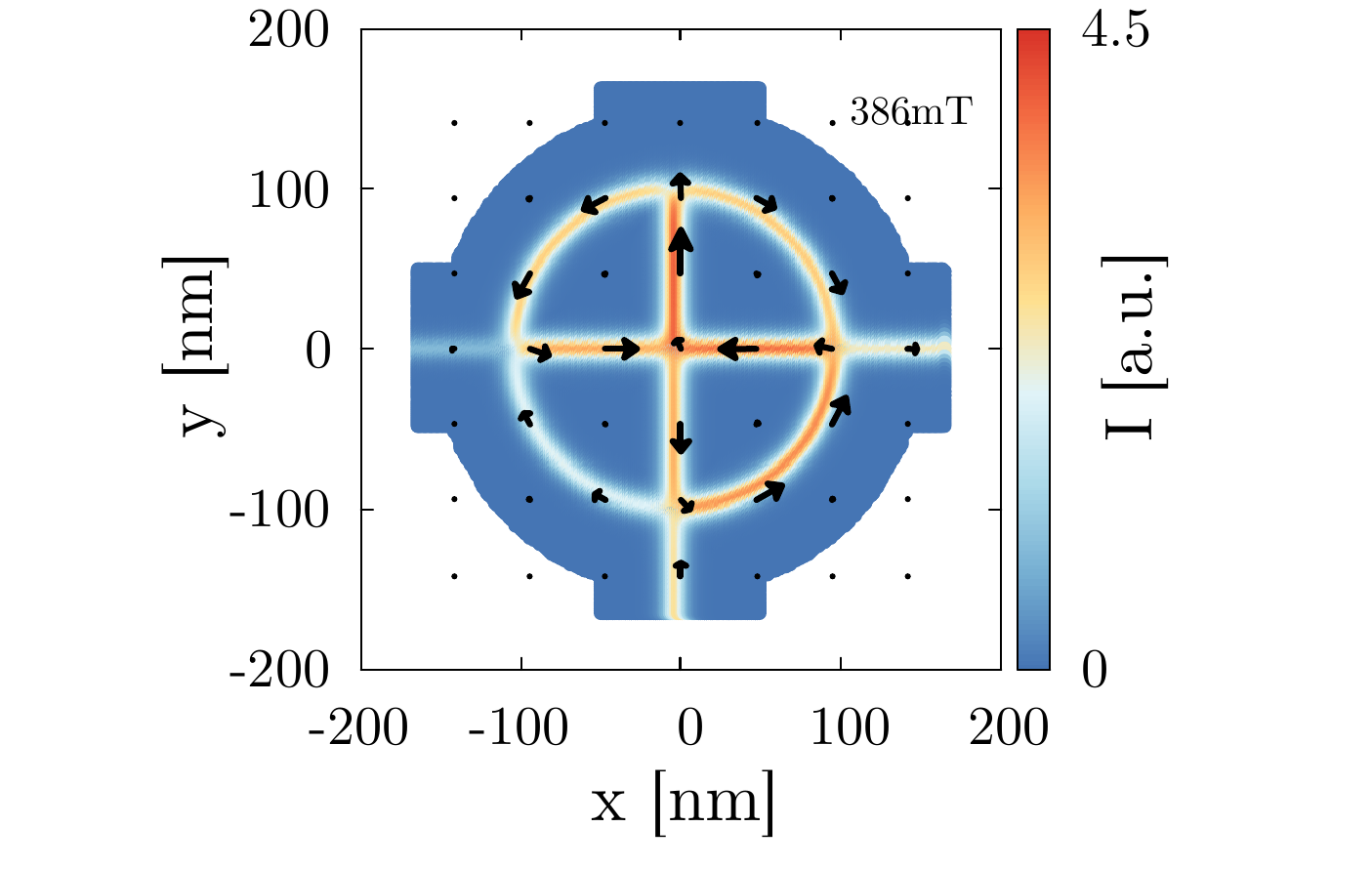}\put(-100,95){(f)}
\caption{ Current distribution maps for different magnetic field magnitudes: (a) 0 mT, (b) 138 mT, (c) 174 mT, (d) 200 mT, (e) 324 mT, and (f) 386 mT for $E_F=20$ meV (see Fig.~\ref{fig:gab_2} for conductance matrix element and resistance).  For each map the color indicates the averaged current amplitude $I$, while the arrows indicate the direction of this current.}
\label{fig:cmap}
\end{figure}

 The current circulation paths for $E_F=20$ meV are presented in Fig. \ref{fig:cmap}. For magnetic fields such that $A=B$ 
Fig. \ref{fig:gab_2}(a),
 the current is evenly distributed from terminal 1 to the left and the right leads  (Fig. \ref{fig:cmap}(a,c)), while for stationary points $0=\frac{\partial A }{\partial \mathcal{B}}=\frac{\partial B }{\partial \mathcal{B}}$ (Fig. \ref{fig:cmap}(b,d,e,f)) one can distinguish current loops around quarters of the ring.

\begin{figure}[htbp]
\centering
\includegraphics[scale=0.27,trim=17 60 0 0 ,clip]{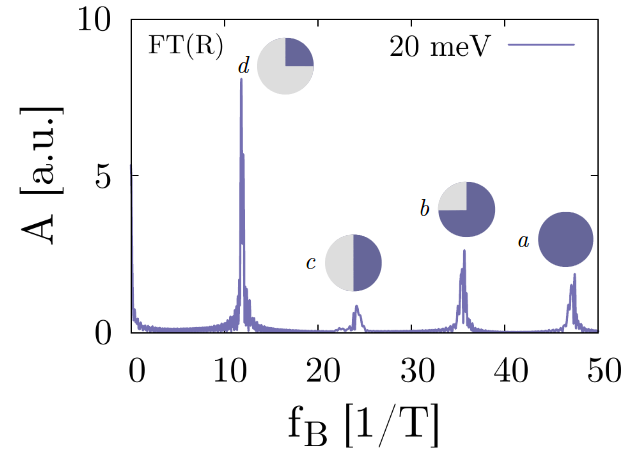}
\includegraphics[scale=0.27,trim=71 60 0 0 ,clip]{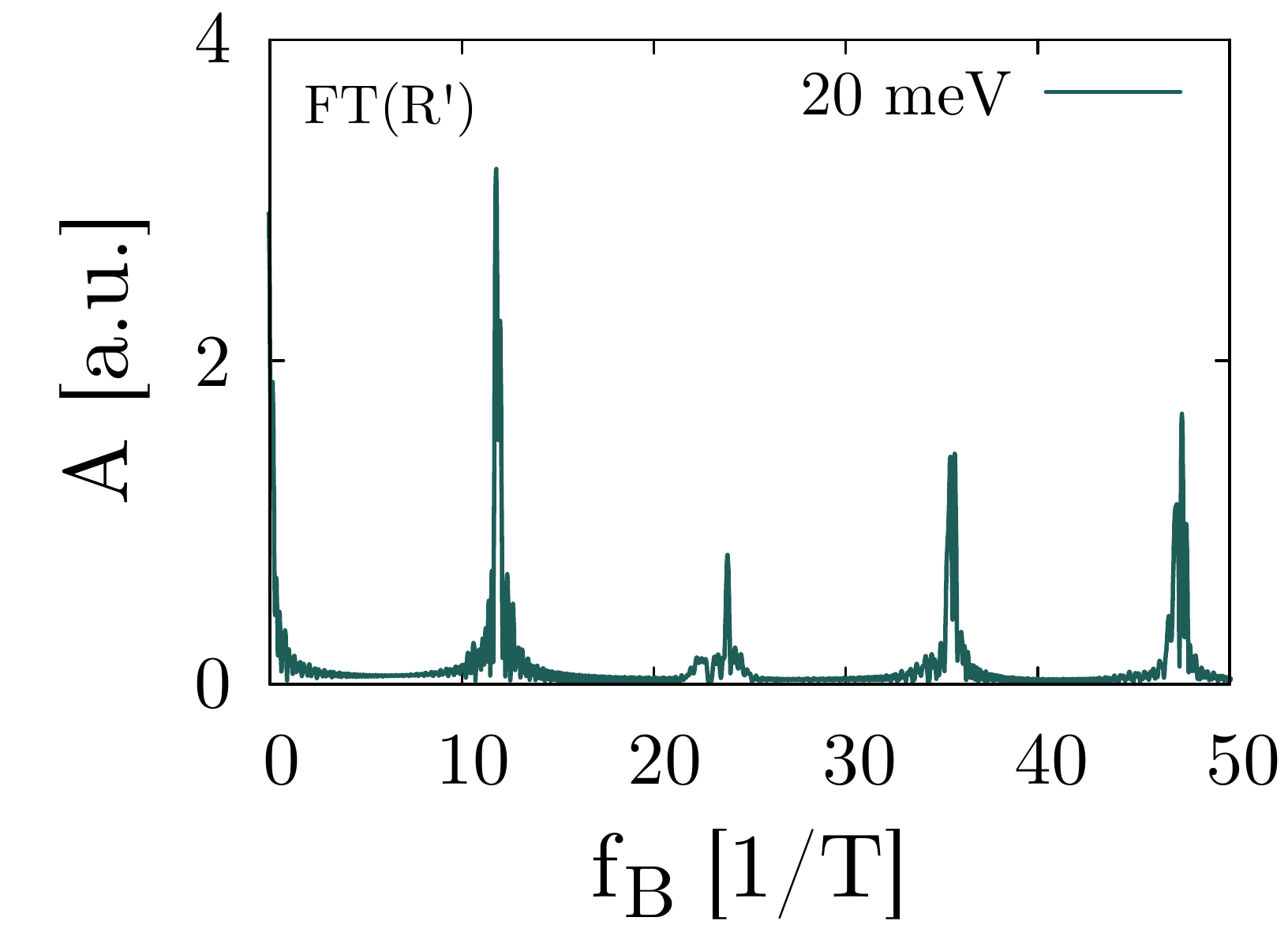}

\includegraphics[scale=0.27,trim=17 0 0 0 ,clip]{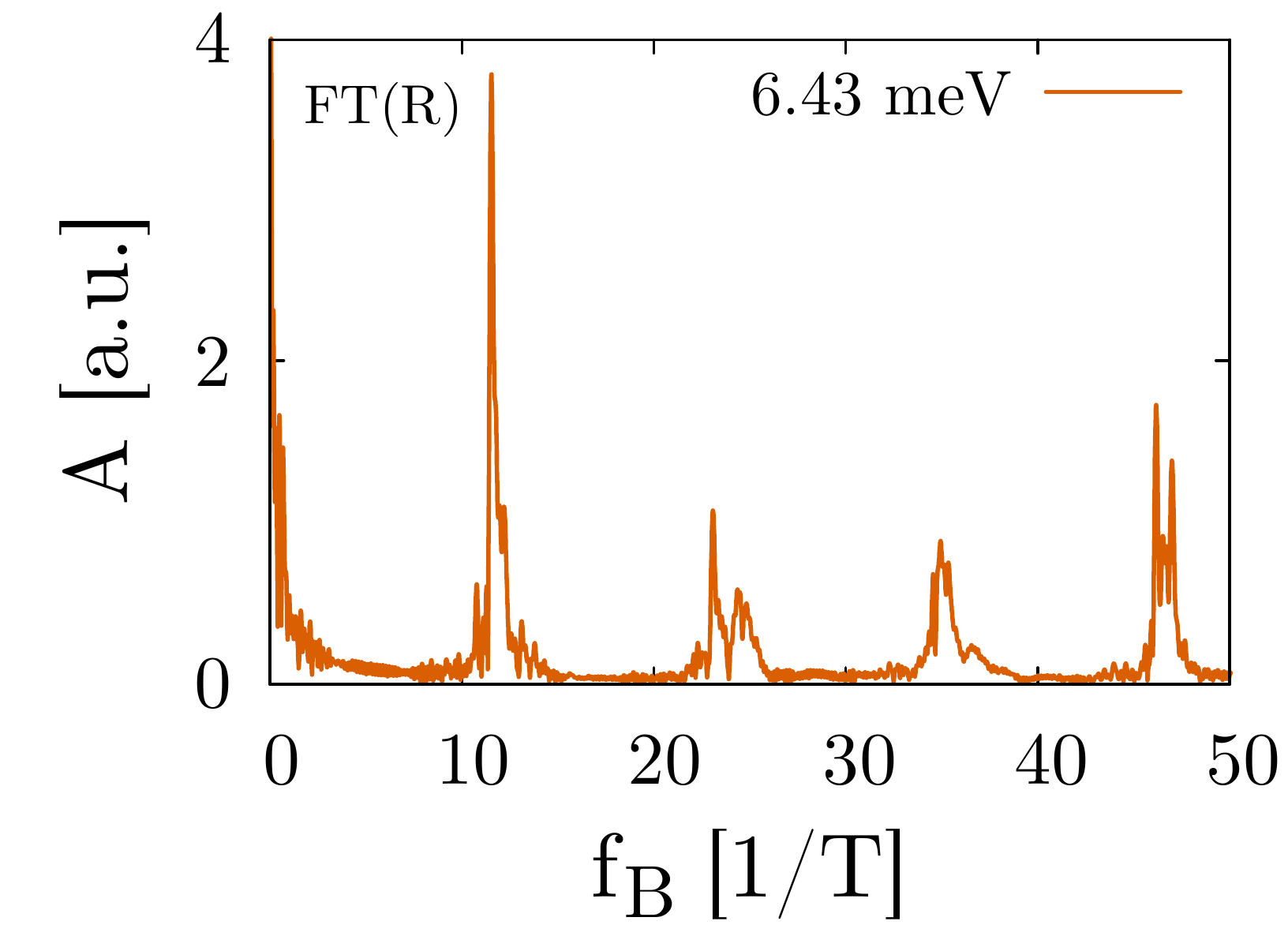}
\includegraphics[scale=0.27,trim=71 0 0 0 ,clip]{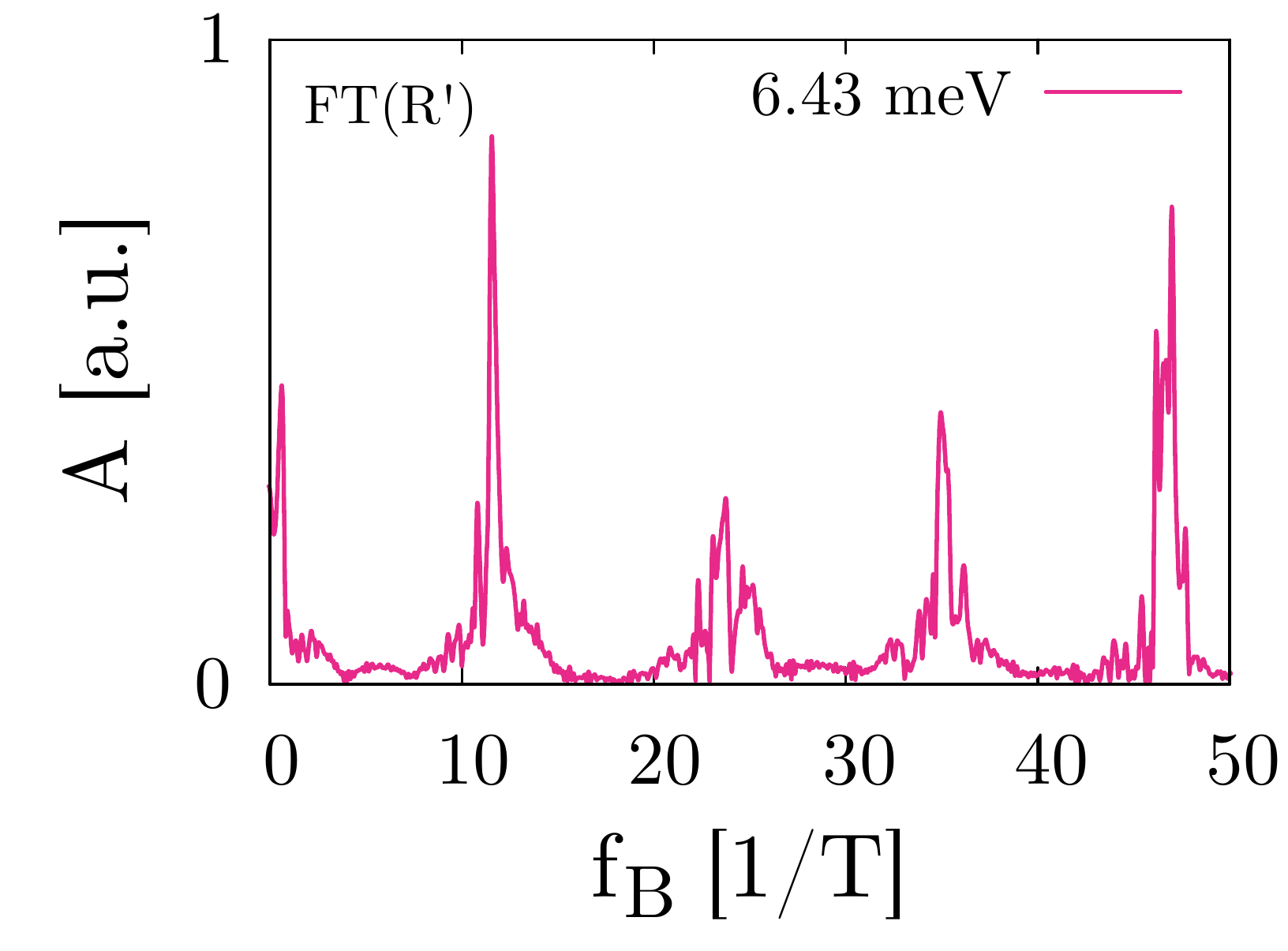}
\caption{ Fourier transform of $R(\mathcal{B})$ and $R'(\mathcal{B})$ data for $E_F$ 20 meV and 6.43 meV. Input magnetic field range $\mathcal{B}$ was set to [0:40] T. Inset icons indicate the area encircled by the currents to produce Aharonov-Bohm periodicity
corresponding to the peak  with $T_a$ = 132mT for the entire ring area, $\frac{3}{4}$ ring with $T_b$ = 176mT, half of the ring with $T_c$ = 264mT and quarter of the ring with $T_d$ = 528mT.   }
\label{fig:fourier}
\end{figure}

By taking the Fourier transform of the resistance $R$ and $R'$ (from Fig.~\ref{fig:gab_2}(b) and \ref{fig:gab_642}(b), respectively) for magnetic field $\mathcal{B}$ range from 0 to 40 T we can distinguish 4 characteristic peaks (Fig.~\ref{fig:fourier}) $f_\mathcal{B} = \{11.9, 23.8, 35.7, 47.6\} \frac{1}{T}$ associated to periods ($\Delta \mathcal{B} = 2\pi/f_\mathcal{B} $) $\Delta \mathcal{B} = \{\text{528 mT, 264 mT, 176 mT, 132 mT}\},$ respectively. For each period the area $\Lambda$ can be calculated as $\Lambda=\pi R^2$, and using the Aharonov-Bohm formula for period $\Delta \mathcal{B} = \frac{h}{e\Lambda}$ we obtain
\begin{equation}
\Lambda = \frac{h}{e\Delta \mathcal{B}}.
\label{eq:ab_area}
\end{equation}
In our calculations the channel ring has radius $\mathcal{R} = 100$ nm and area $\Lambda_0 = \pi \mathcal{R}^2$, hence
\begin{equation}
\mathbf{\Lambda}=\frac{\Lambda}{\Lambda_0} = \frac{h}{ e\Delta \mathcal{B} \pi \mathcal{R}^2}
\end{equation}
is the fraction of the ring area responsible for the Aharonov-Bohm interference. Thus, taking the $\{\Delta \mathcal{B}\}$ list from the Fourier transform we obtain $$ \mathbf{\Lambda} = \left\{\frac{1}{4},\frac{1}{2},\frac{3}{4},1\right\}, $$ for peaks 1 -- 4 from left to right in Fig. \ref{fig:fourier}, respectively.
The leftmost peak that corresponds to the interference paths that encircle a quarter of the ring is the most pronounced. 
In the current distribution in Fig.~\ref{fig:cmap} one can indicate
the paths that encircle a few quarters of the ring, but the fundamental period 
corresponds to 1/4 of the ring.


\subsection{Larger ring, nonchiral bands, weaker vertical field}

The clear Aharonov-Bohm oscillations of the resistance presented above were obtained for a system with a relatively small radius, narrow flip length and a very strong vertical electric field with only chiral bands at the Fermi level. 

\begin{figure}[htbp]
\centering
\includegraphics[width=0.20\textwidth, trim=0 0 0 0 ,clip]{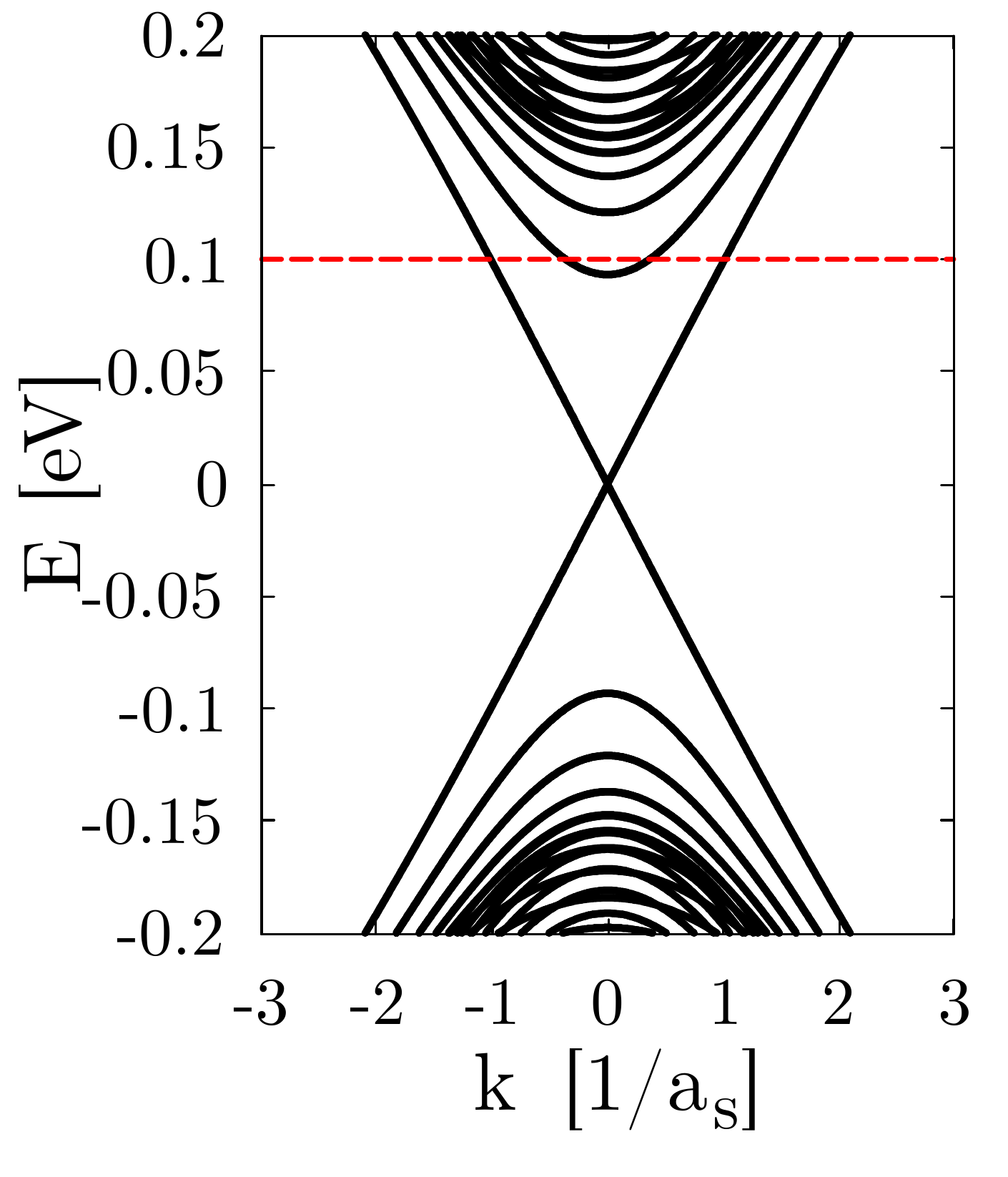}
\includegraphics[width=0.20\textwidth, trim=0 0 0 0 ,clip]{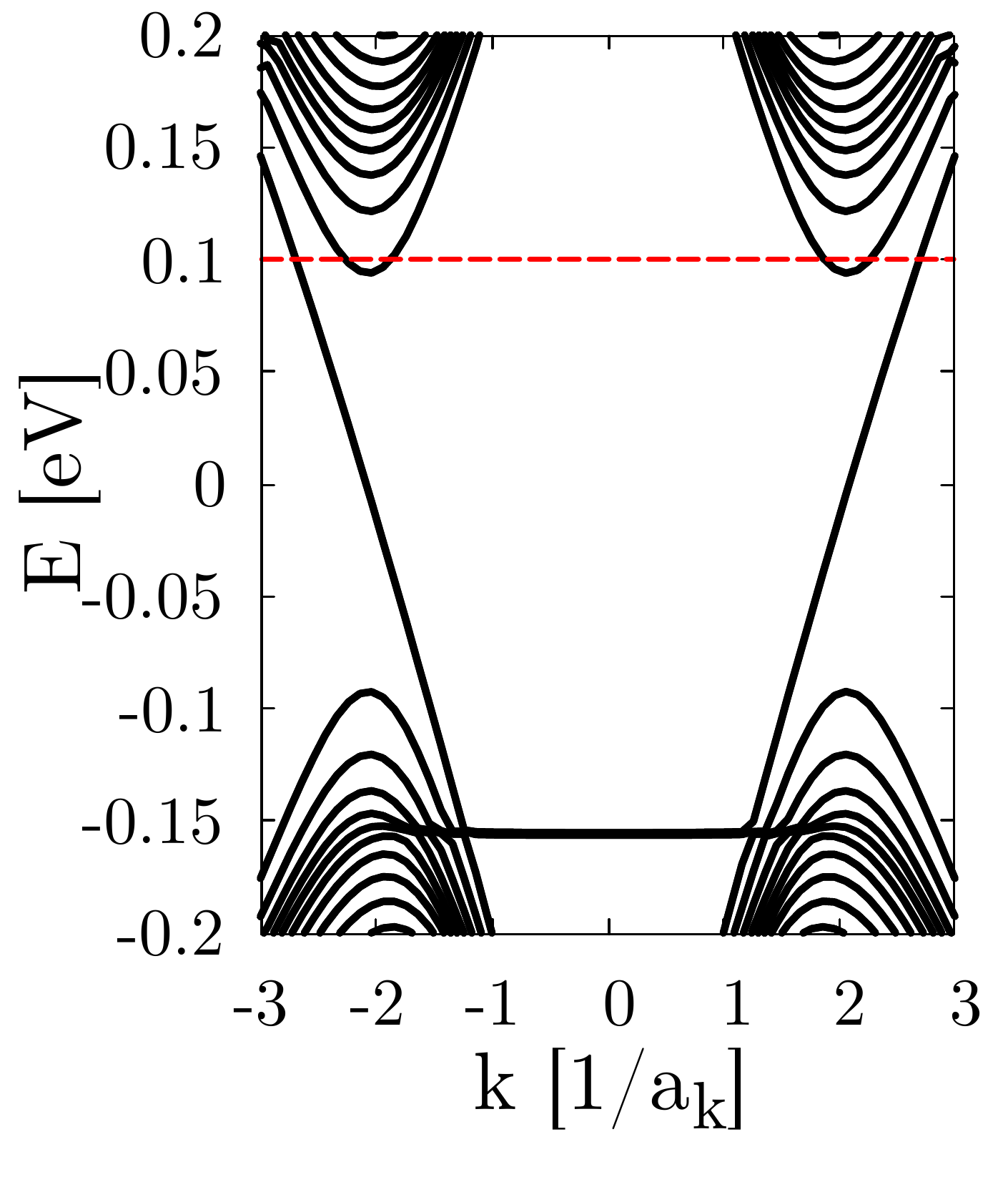}
\put(-210,110){(a)}\put(-105,110){(b)}
\caption{ (a) Band structure for the armchair-type lead number 1 and (b) for the zigzag-type lead number 2 with $V_G=200$ meV and $\lambda=12$ nm. Red line denotes $E_F=100$ meV that includes one chiral and one (two) non-chiral subbands in armchair (zigzag) lead. $a_s = 2\sqrt{3}a$ is the translation vector of the supercell of the armchair nanoribbon.}
\label{fig:sil_200_band}
\end{figure}

\begin{figure}[htbp]
\centering
\includegraphics[width=0.48\textwidth, trim=0 0 0 0 ,clip]{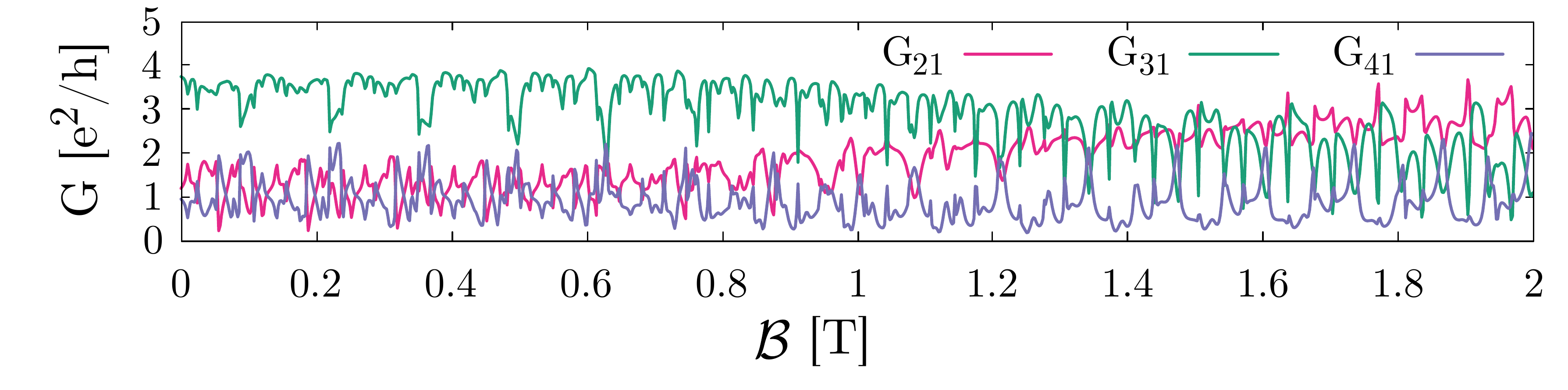}\put(-250,60){(a)}\\
\includegraphics[width=0.48\textwidth, trim=0 0 0 0 ,clip]{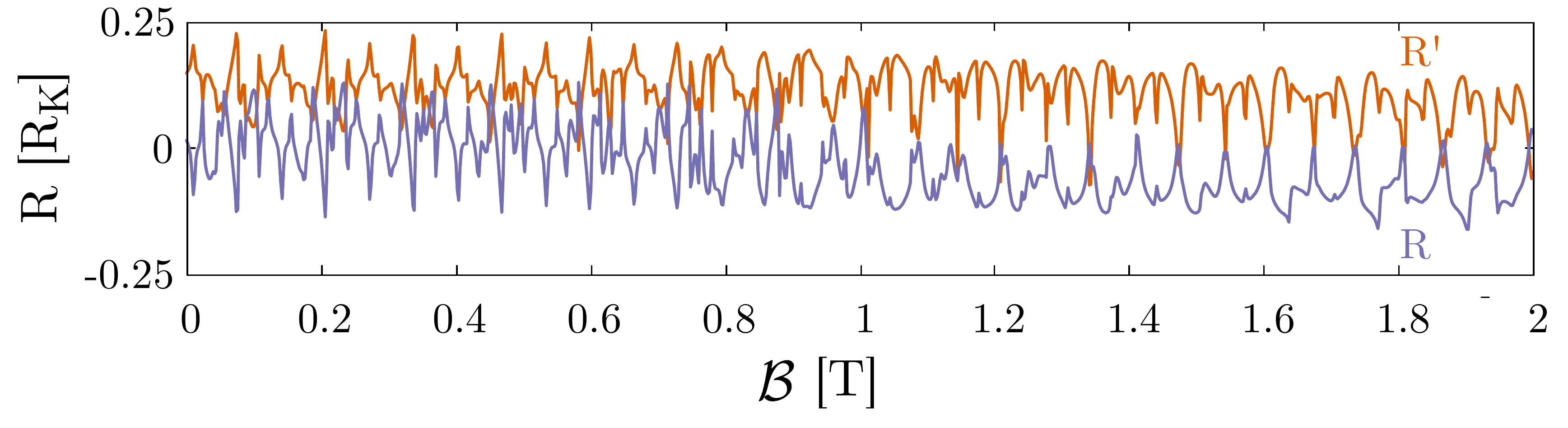}\put(-250,60){(b)}
\caption{ (a) Conductance matrix elements $G_{21},G_{31}, G_{41}$ and (b) resistances $R,R'$ calculated from the inverted conductance matrix Eq. (\ref{eq:cm}) for $E_F=100$ meV and parameters $V_G=200$ meV and $\lambda=12$ nm. }
\label{fig:sil_200_nonchiral}
\end{figure}

Let us consider a system with larger field inversion length increased from $\lambda=4$ nm to $12$ nm and $E_F=100$ meV, for which nonchiral modes appear at the Fermi level (Fig. \ref{fig:sil_200_band}). 
The results for the conductance matrix elements and the resistance are plotted in Fig. \ref{fig:sil_200_nonchiral}(a).  The non-chiral currents transfer across the ring from lead 1 to 3, 
see  $G_{31}\neq 0$ (Fig. \ref{fig:sil_200_nonchiral}(a)),
which is forbidden for the chiral bands. 
In presence of the nonchiral bands the conductance matrix has no longer the form given by Eq. (\ref{eq:simply})
and a general formula needs to be applied to calculate the resistances $R$ and $R'$ by inverting the conductance matrix. The results of  $R$ and $R'$ calculations are presented in Fig. \ref{fig:sil_200_nonchiral}(b). Matching peaks (dips) to a periodic pattern we observe periodicity 
of the $R(\mathcal{B})$ plot with mean spacing $\Delta \mathcal{B}$ of 66 mT and 131 mT, which for $\mathcal{R}=200$ nm correspond to flux quantum threading 1/2 and 1/4 of the ring area, respectively. In the presence of the nonchiral bands, the amplitude of the $R'$ oscillations becomes comparable to the ones of $R$. 

\begin{figure}[htbp]
\centering
\includegraphics[width=0.48\textwidth, trim=0 0 0 0 ,clip]{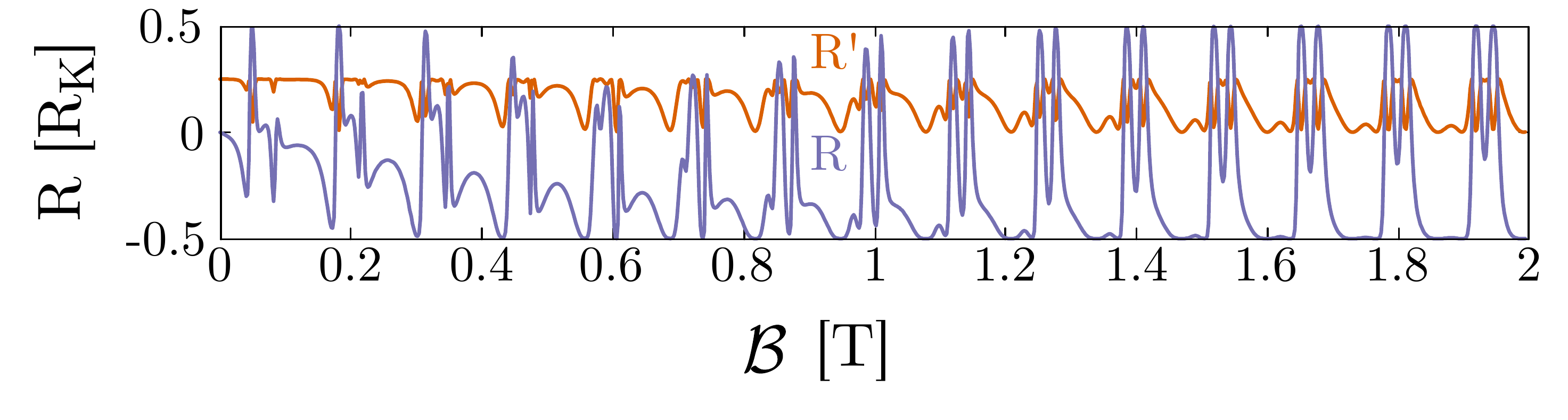}
\caption{ Resistances $R,R'$ for $V_G=200$ meV and inversion length  $\lambda=12$ nm at $E_F=20$ meV.}
\label{fig:sil_200_12nm}
\end{figure}

For the same parameters $\mathcal{R}=200$ nm and $\lambda=12$ nm but a
lower Fermi level $E_F=20$ meV (see Fig. \ref{fig:sil_200_band}) we reproduce the regular oscillations (Fig. \ref{fig:sil_200_12nm}) of the purely chiral case presented above for $\mathcal{R}=100$ nm and $\lambda=4$ nm. 

\begin{figure}[htbp]
\centering
\includegraphics[width=0.48\textwidth, trim=0 0 0 0 ,clip]{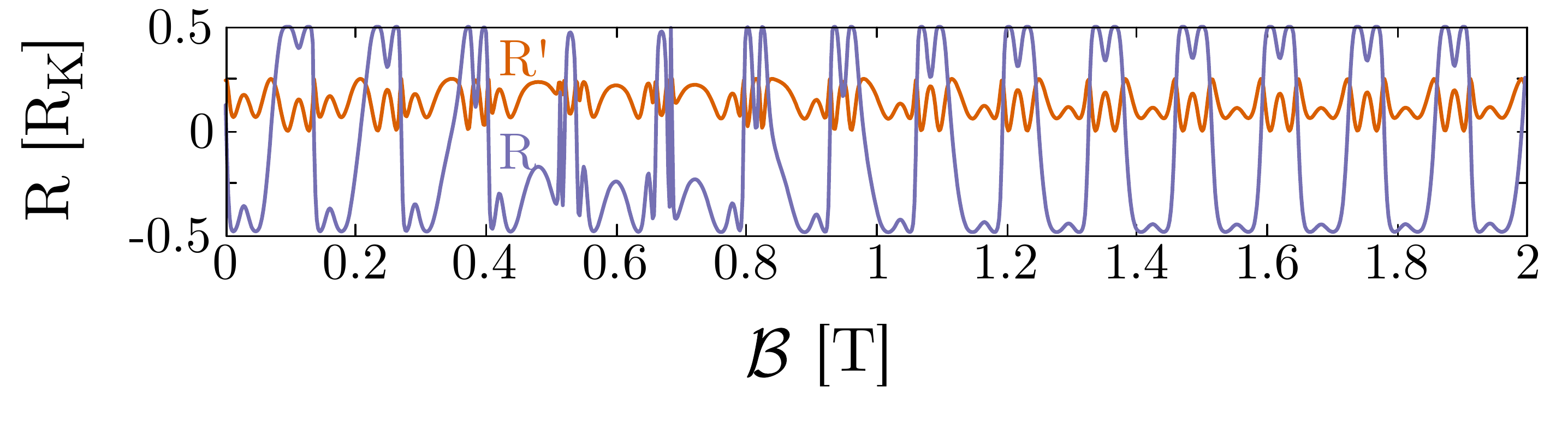}
\caption{ $R$ and $R'$ resistances for 10 times weaker gate potential $V_G=20$ meV with $\lambda=4$ nm at $E_F=2$ meV. }
\label{fig:sil_200_weak}
\end{figure}

The assumed potential of $ \pm 0.2$ eV at each sublattice of the buckled silicene requires a giant vertical field of  the order of 10 V/nm.  The vertical field applied for two-dimensional crystals can be very large without inducing the breakdown due to atomic width of the system. However, the
fields considered for the silicene \cite{ni} and the ones applied to bilayer-graphene \cite{bigf1,bigf2}  are of the order of 1 V/nm only. In order to verify that the  effects described above
can be observed for similar electric fields we performed calculations for
10 times weaker gate potential $V_G=20$ meV at $E_F=2$ meV for  $\mathcal{R}=200$ nm and $\lambda=4$ nm  with only the single linear chiral mode  at the Fermi level. The Aharonov-Bohm oscillations can be resolved in $R$ and $R'$ dependence on the magnetic field (Fig. \ref{fig:sil_200_weak}), with the larger visibility of the $R$ oscillations as above.

\begin{figure}[htbp]
\centering
\includegraphics[width=0.3\textwidth, trim=30 30 0 0 ,clip]{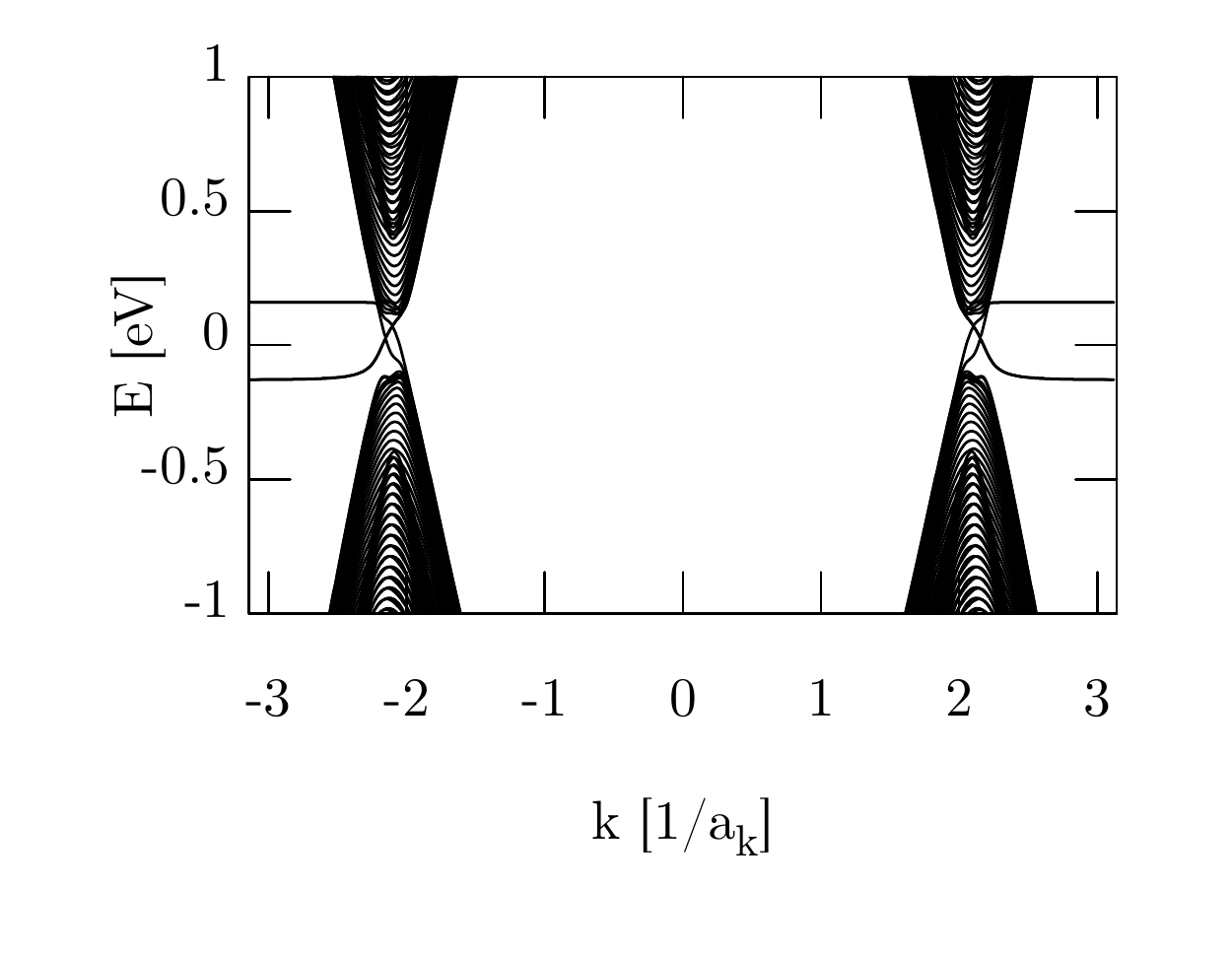}
\put(-160,100){(a)}
\\
\includegraphics[width=0.15\textwidth, trim=30 30 0 0 ,clip]{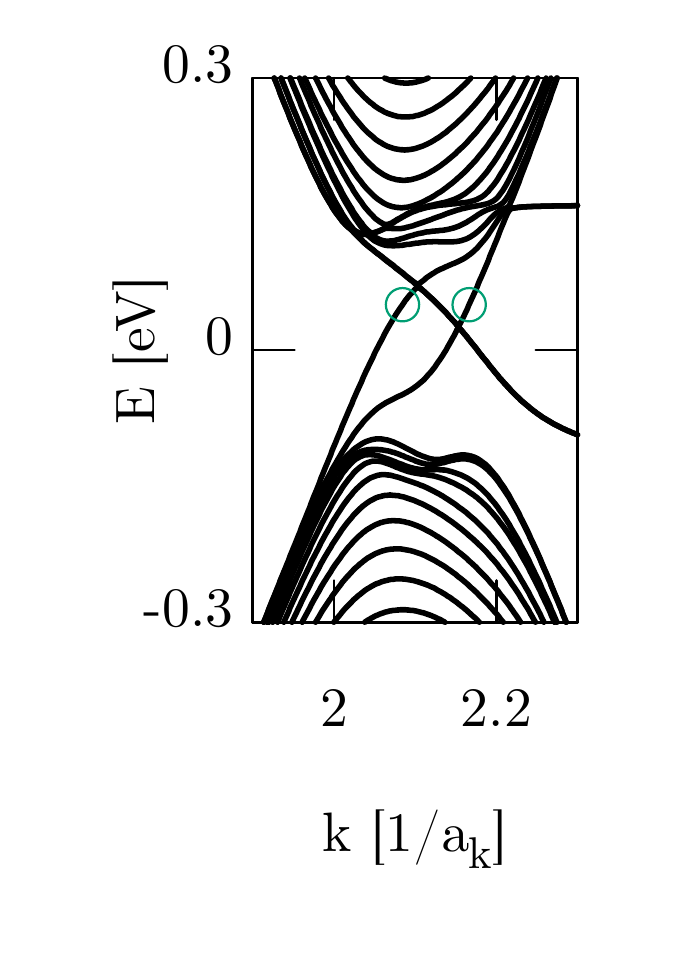}
\includegraphics[width=0.15\textwidth, trim=30 30 0 0 ,clip]{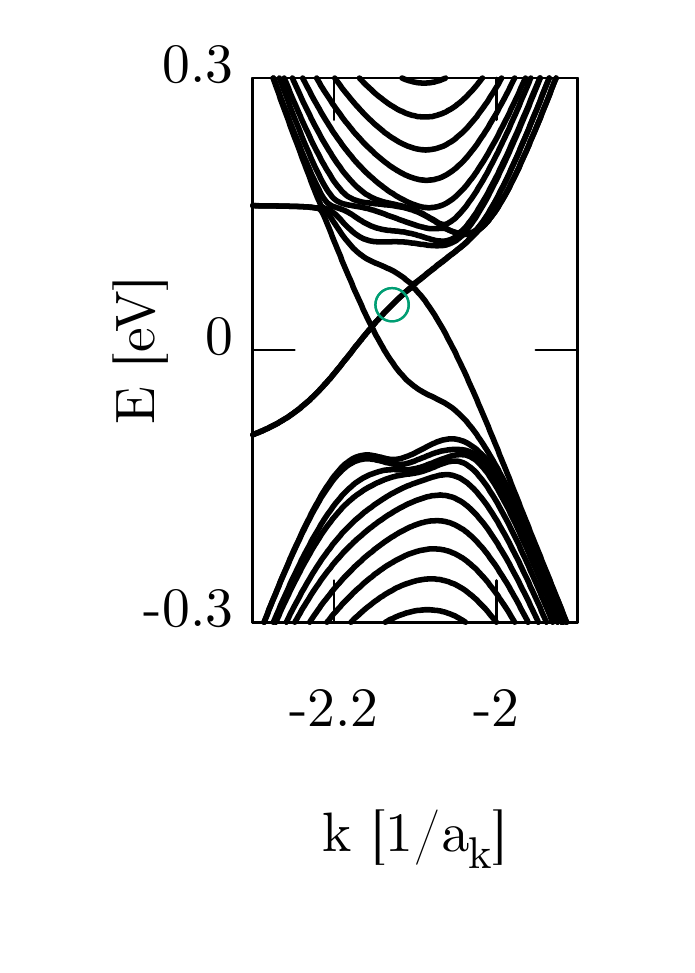}
\put(-160,100){(b)}\put(-82,100){(c)}\put(-5,100){(d)}
\includegraphics[width=0.15\textwidth, trim=30 30 0 0 ,clip]{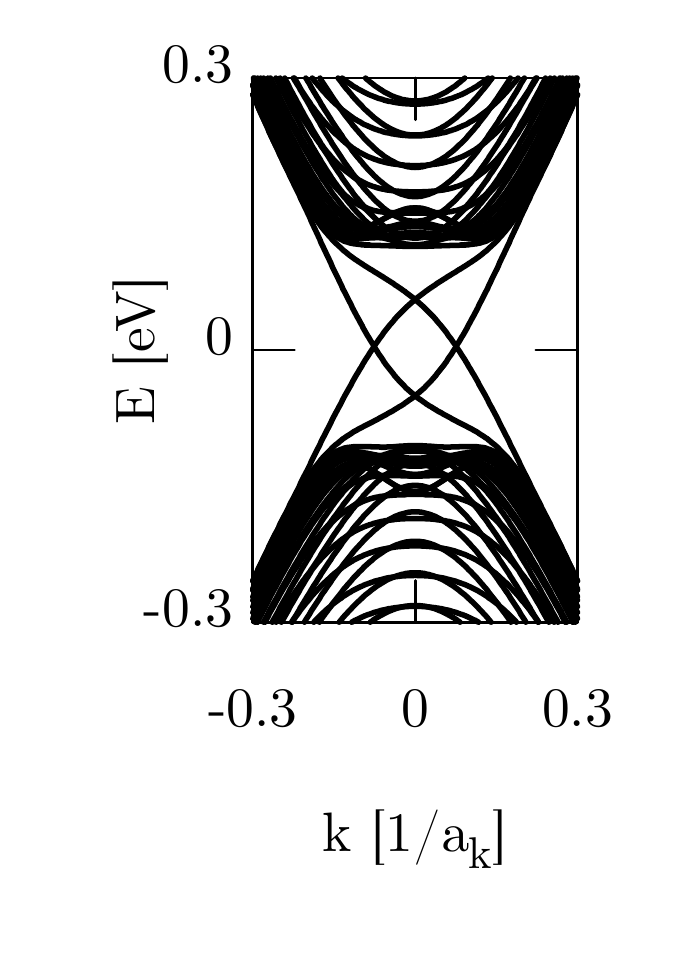}
\caption{ (a) Band structure for lead 1 (with zigzag edges) (b) and its zoom in the vicinity of $K'$ valley, and (c) in the vicinity of $K$ valley, and (d) for lead 2 (with armchair edges) with electric field defined as described in Fig. \ref{fig:sch1}. The green circles in (b) and (c) indicate the modes outgoing from lead 1.}
\label{fig:bnd_blg}
\end{figure}

\begin{figure}[htbp]
\centering

\includegraphics[scale=0.29, trim=10 60 0 0 ,clip]{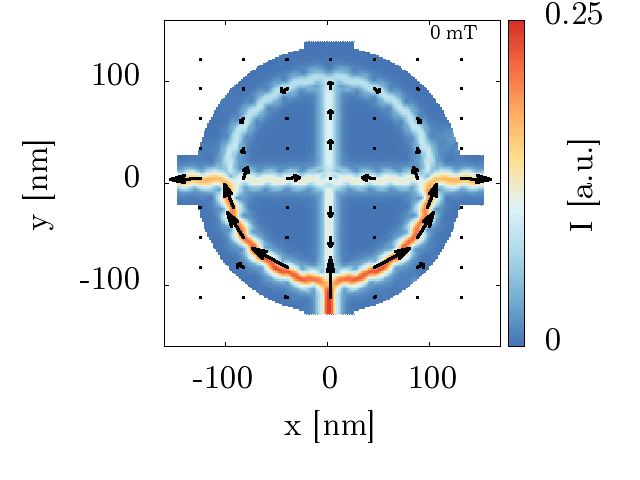} \put(-110,85){(a)}
\includegraphics[scale=0.29, trim=120 60 0 0 ,clip]{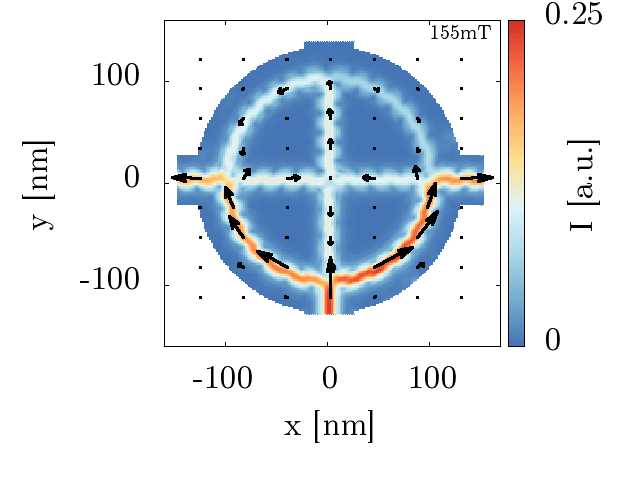} \put(-110,85){(b)}

\includegraphics[scale=0.29,trim=10 0 0 0 ,clip]{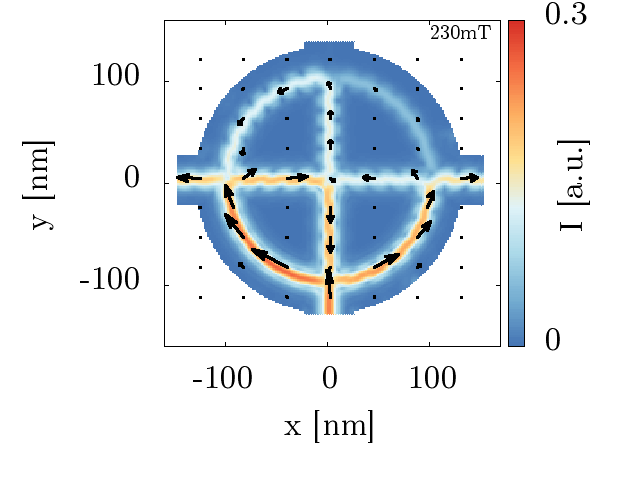} \put(-110,105){(c)}
\includegraphics[scale=0.29,trim=120 0 0 0 ,clip]{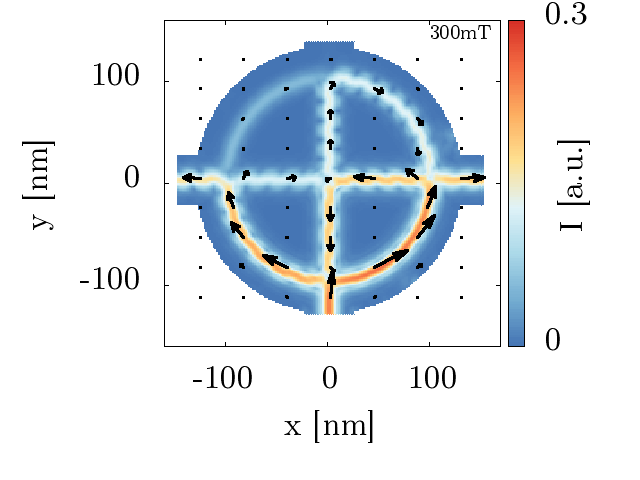} \put(-110,105){(d)}
\caption{ Current distribution maps in bilayer graphene ring at $E_F=50$ meV for different magnetic field magnitudes: (a) 0 mT, (b) 155 mT, (c) 230 mT, and (d) 300 mT. For each map the color indicates the averaged current amplitude $I$, and black arrows present the direction of this current.}
\label{fig:blg_cmap}
\end{figure}

\begin{figure}[htbp]
\centering
\includegraphics[width=0.96\columnwidth]{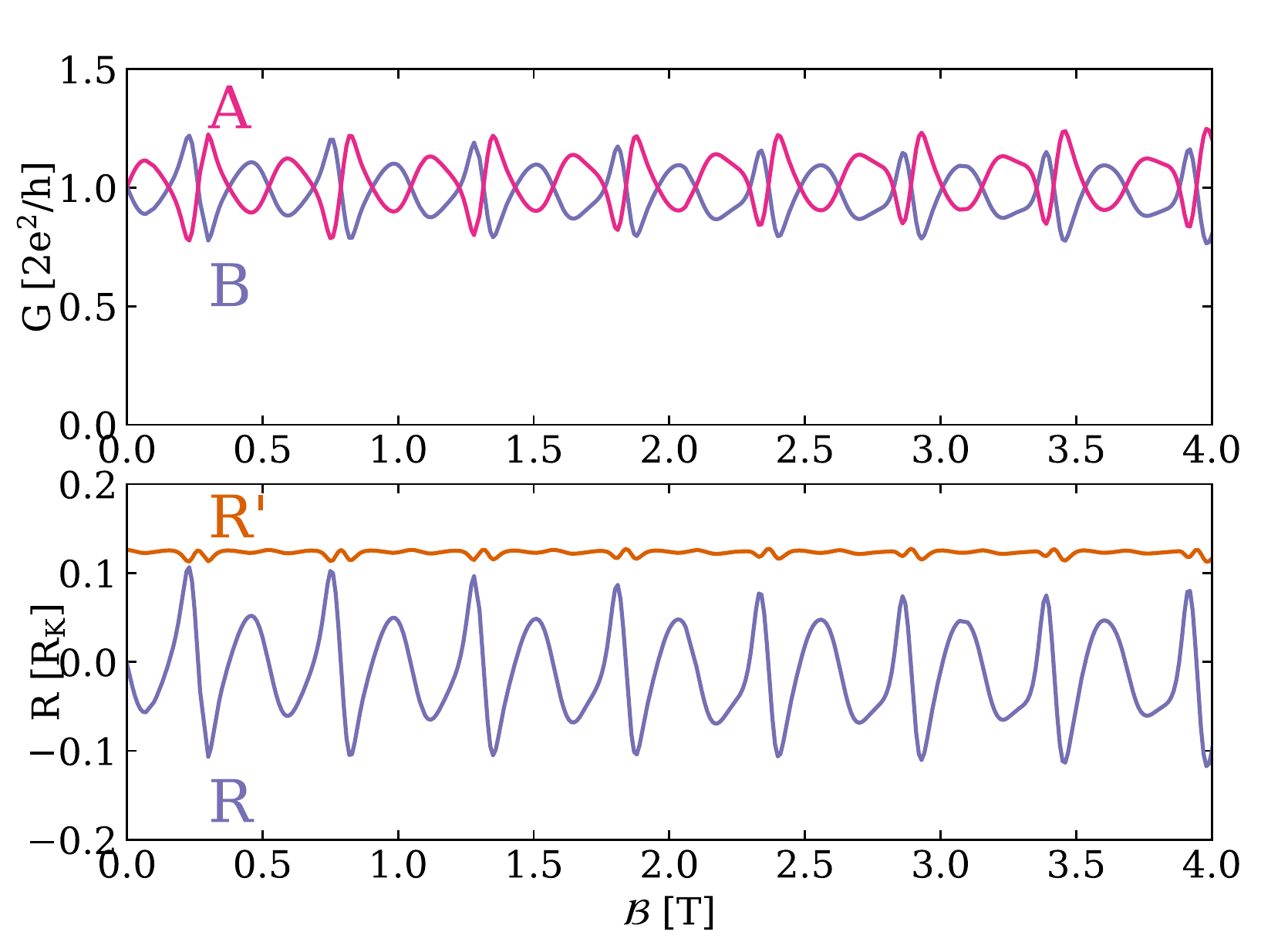}\put(-240,165){(a)} \put(-240,85){(b)}
\caption{  (a) Conductance plot for simplified conductance matrix elements (Eq. \ref{eq:simply}) for the bilayer graphene system in external magnetic field at fixed Fermi level $E_F$ = 50 meV. (b) Resistance $R'$ for the case in Fig.\ref{fig:conf}(a) and $R$ [case Fig.\ref{fig:conf}(b)] in units of the von Klitzing constant $R_K$. }
\label{fig:blg_gab}
\end{figure}

\begin{figure}[htbp]
\centering
\includegraphics[width=\columnwidth]{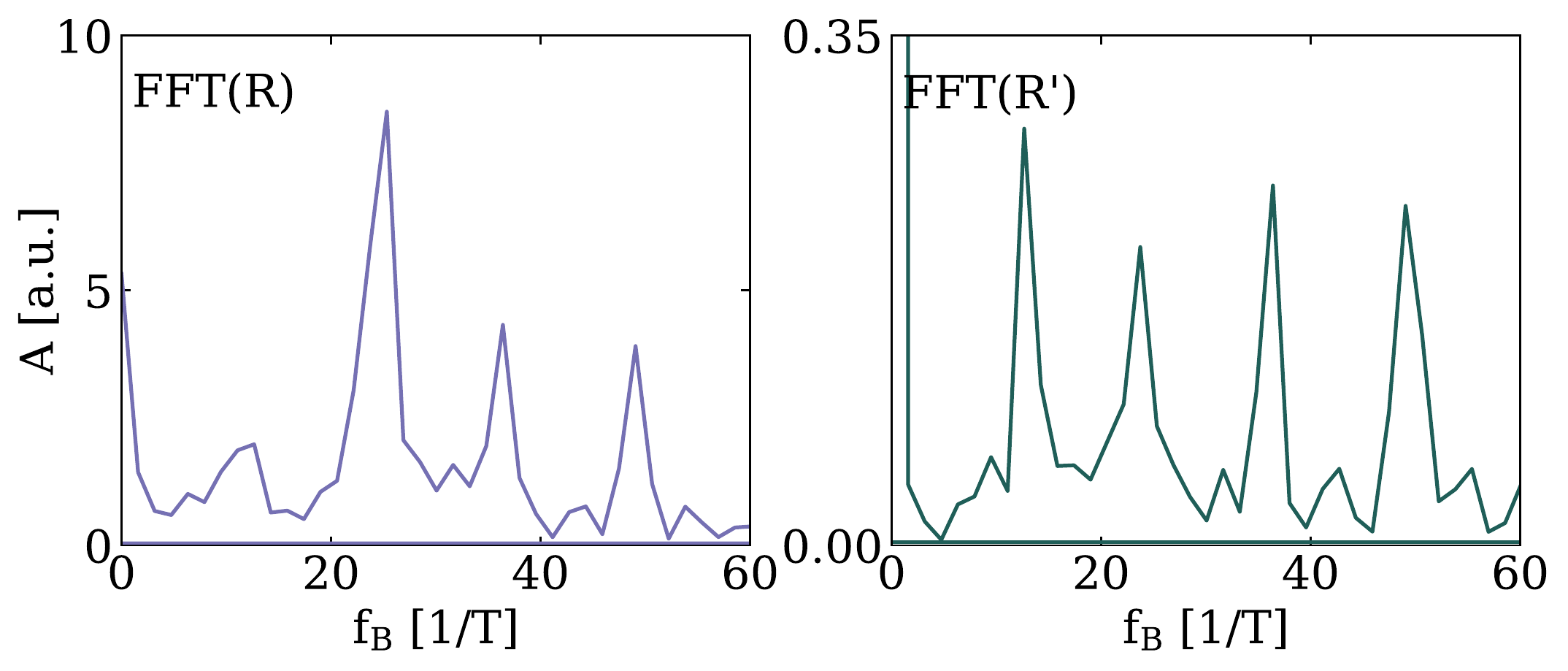}\put(-245,105){(a)} \put(-125,105){(b)}

\caption{ Fourier transform of (a) the $R$ and (b) the $R'$ signal of Fig.~\ref{fig:blg_gab} in the bilayer graphene system. }
\label{fig:blg_fft}
\end{figure}
\section{Bilayer-graphene-based system}
For bilayer graphene system the results are qualitatively similar as for silicene, with the difference that for the Fermi energy within the energy gap we have max$(G_{\xi\eta})=4G_0$, as the number of topological states is doubled due to the presence of two layers. This can be seen in the band structure of the armchair input leads in Fig.~\ref{fig:bnd_blg}(d). For the zigzag leads, Fig.~\ref{fig:bnd_blg}(a, b, c), within the energy gap the edge states occur that, however, do not contribute to the inter-lead conductance. $G_{11}$ and $G_{33}$ is always equal to 2, with the edge modes being completely backscattered, and only the flip-modes leaving the zigzag leads.

Fig.~\ref{fig:blg_cmap} shows the current distribution in the bilayer graphene system for $E_F=50 $ meV
for the electron incident from the lower lead.  As in silicene, the current cannot pass to the upper lead. Instead, we observe only the transfer to one of the two nearest leads. For $A=B$ [Fig.~\ref{fig:blg_cmap}(a) and (b)] the current is evenly distributed within the system, while at the extrema of $A$ and $B$ [Fig.~\ref{fig:blg_cmap}(c) and (d)]  the current distribution is asymmetric, and loops around quarters of the ring are more pronounced.

The conductance matrix elements in Fig.~\ref{fig:blg_gab}(a), and the resistance  in Fig.~\ref{fig:blg_gab}(b)  manifest oscillations of the  periodicity corresponding to a single or several quarters of the ring as for silicene.  In the Fourier transform of the resistance in Fig.~\ref{fig:blg_fft}(a) and (b) we find peaks at the frequencies
$f_\mathcal{B} = \{12.7, 23.8, 36.4, 49\} \frac{1}{T}$ associated with the periods ($\Delta \mathcal{B} = 2\pi/f_\mathcal{B} $) $\Delta \mathcal{B} = \{\text{495 mT, 263 mT, 173 mT, 128 mT}\},$ respectively. These correspond roughly to the area of one, two, three, or four quarters of the ring, respectively.

\section{Summary and conclusions}
We have studied  Aharonov-Bohm interferometers with chiral channels defined by inversion of the vertical
electric field in silicene and bilayer graphene.  The valley protected channels induced by inhomogeneous electric field in silicene and bilayer graphene in clean conditions i.e. without the backscattering (due to the intervalley transitions)
can serve for the observation of the Aharonov-Bohm oscillations provided that four (instead of two) terminals are attached
to the system.  
The Aharonov-Bohm oscillations of four-probe resistance
with large visibility are observed when a direct electron transfer between terminals (chosen as the current probes) is forbidden.  The fundamental period of the resistance oscillations corresponds to a quarter of the ring, or to the smallest loop that a chiral current encircles within the structure.

\section*{Acknowledgments}
B.R. is supported by Polish government budget for science in 2017-2021 as a research project under the program  "Diamentowy Grant" (Grant No. 0045/DIA/2017/46), by the EU Project POWR.03.02.00-00-I004/16 and NCN grant UMO-2019/32T/ST3/00044. A.M-K. is supported with   "Diamentowy Grant" (Grant No. 0045/DIA/2017/46). The calculations were performed on PL-Grid Infrastructure on Rackserver Zeus at ACK-AGH Cyfronet.

\bibliographystyle{apsrev4-1}
\bibliography{bib_silicene}

\end{document}